\documentclass[twocolumn,twocolappendix]{aastex701}

\usepackage[T1]{fontenc}

\newsavebox{\foobox}
\newcommand{\slantbox}[2][0]{\mbox{%
        \sbox{\foobox}{#2}%
        \hskip\wd\foobox
        \pdfsave
        \pdfsetmatrix{1 0 #1 1}%
        \llap{\usebox{\foobox}}%
        \pdfrestore
}}
\newcommand\unslant[2][-.25]{\slantbox[#1]{$#2$}}

\begin{document}

\title{The Effect of External Photoevaporation on the Disk Fraction in M17}

\correspondingauthor{Samuel Millstone}
\email[show]{samuelmillstone@gmail.com}

\author[0000-0003-3512-5885]{Samuel Millstone}
\affiliation{Department of Physics and Astronomy, Rice University, Houston, TX 77005, USA}
\email{sm270@rice.edu}
\email{samuelmillstone@gmail.com}

\author[0000-0002-3887-6185]{Megan Reiter}
\affiliation{Department of Physics and Astronomy, Rice University, Houston, TX 77005, USA}
\email{Megan.Reiter@rice.edu}

\author[0000-0002-5306-4089]{Morten Andersen}
\affiliation{European Southern Observatory, 85748 Garching bei München, Germany}
\email{Morten.Andersen@eso.org}

\author[0000-0002-9593-7618]{Thomas J. Haworth}
\affiliation{School of Physical and Chemical Sciences, Queen Mary University of London, London, E1 4NS, UK}
\email{t.haworth@qmul.ac.uk}

\author[0000-0002-9051-1781]{Dominika Itrich}
\affiliation{Steward Observatory, University of Arizona, 933 North Cherry Avenue, Tucson, AZ 85721, USA}
\email{ditrich@arizona.edu}

\author[0000-0002-5456-523X]{Anna McLeod}
\affiliation{Centre for Extragalactic Astronomy, Department of Physics, Durham University, Durham, DH1 3LE, UK}
\email{anna.mcleod@durham.ac.uk}

\author[0000-0002-1474-7848]{Richard J. Parker}
\affiliation{Astrophysics Research Cluster, School of Mathematical and Physical Sciences, The University of Sheffield, Hicks Building, Sheffield, S3 7RH, UK}
\email{r.parker@sheffield.ac.uk}

\author[0000-0002-7501-9801]{Andrew Winter}
\affiliation{School of Physical and Chemical Sciences, Queen Mary University of London, London, E1 4NS, UK}
\email{Andrew.winter@qmul.ac.uk}

\author[0000-0002-6091-7924]{Peter Zeidler}
\affiliation{AURA for the European Space Agency, ESA Office, STScI, 3700 San Martin Drive, Baltimore, MD 21218, USA}
\email{zeidler@stsci.edu}

\begin{abstract}

\noindent A major obstacle to improving models of planet formation is understanding how the local environment influences the lifetime of the disks in which they form. The spread in observed disk lifetimes is caused by effects both observational (e.g., target selection, survey sensitivity) and physical (e.g., disk destruction by internal and external photoevaporation); however, the degree to which each plays a role remains poorly constrained. Isolating the impact of external photoevaporation on the disk lifetime benefits from the inclusion of low-mass ($\lesssim0.5$~M$_{\odot}$) YSOs, for which this effect is most predominant. In this work, we measure the inner disk fraction from JHK excess in the $\sim\!6000$~M$_{\odot}$, $\sim$1~Myr-old star-forming region M17. Using VLT/HAWK-I, we perform a deep photometric survey of an $\sim$$8^{\prime}\times8^{\prime}$ field towards the region. The $\sim$$4$ times greater sensitivity and $\sim$$2-3$ times higher resolution than previous surveys of M17 reveal 10,339~sources. We select cluster members using the Massive Young Star-Forming Complex Study in Infrared and X-ray (MYStIX) catalog and find a disk fraction of $28\pm2\%$: the first X-ray-selected disk fraction measurement in M17 to include low-mass YSOs, and only the second such measurement in any high-mass star-forming region. After correcting for observational biases, we find no correlation between disk fraction and incident UV flux within M17, likely due to dynamical mixing within the region. However, when compared to other regions of similar age, we find lower disk fractions in regions with higher UV fields, suggesting that external photoevaporation decreases the average disk lifetime.

\end{abstract}

\keywords{\uat{Young Stellar Objects}{1834} --- \uat{Protostars}{1302} --- \uat{Protoplanetary Disks}{1300} --- \uat{Circumstellar Disks}{235} --- \uat{Star Clusters}{1567} --- \uat{Interstellar Radiation Field}{852} --- \uat{Infrared Dark Clouds}{787} --- \uat{Star Formation}{1569} --- \uat{Star Forming Regions}{1565}}


\section{Introduction} \label{sec:intro}
Most young stellar objects (YSOs) form in complex environments of dense gas and dust, surrounded by tens to hundreds of neighboring YSOs \citep[e.g.,][]{La+91, Br+10, Kr+19}. As YSOs evolve, material from their surrounding envelopes of gas and dust collapses and flattens into circumstellar disks. It is in these disks that nearly all planets form \citep[e.g.,][and references therein]{Dr+23}.

Dusty disk material absorbs stellar light from the host star and reemits it in the infrared (IR) beyond $\sim$$2~\unslant\mu$m, \citep{Ca+92, Me+97, Hi+98, Fe+07}, in excess of what is expected from the photosphere of the YSO \citep[e.g.,][]{Me68, RSS76, Ru85}, with excess at shorter wavelengths tracing hotter regions of the disk. Observations show an exponential decrease in the disk fraction as a function of star-forming region (SFR) age, suggesting an average disk lifetime in SFRs within the nearest few kpc of between $1-10$~Myr \citep[e.g.,][]{Ha+01, Ri+18, Pf+22}, though the disk lifetime can be influenced by environmental effects \citep[see e.g.,][]{St+04, Gu+07, Pr1+11, Pr2+11, Gu+16, Co+22, Co+23}.

For example, UV radiation from OB stars strips material from disks around nearby YSOs through external photoevaporation \citep[e.g.,][for two recent reviews]{Ch+87, Jo+98, HA98, WH22, PPF25}. The UV radiation excites and heats the disk material, accelerating it above the escape velocity of the host YSO, evaporating and destroying the disk from the outside in. External photoevaporation is stronger for disks around lower-mass stars, as the hosts have a weaker gravitational hold on their circumstellar disks, so the escape velocity at any given radius is lower \citep[e.g.,][]{Ad+04, Ha+18}.

External photoevaporation has been observed directly for decades in the ``proplyd'' population of the Orion Nebular Cluster \citep[ONC;][]{Od+93, Ri+08}. \cite{Ar+24} examined 12 of these objects with the VLT/MUSE IFU, studying emission line and ionization front morphology to investigate the impact of external photoevaporation. Proplyds are small \citep[$\sim$200~AU;][]{WH22} and thus require a high resolution to observe directly, though spatially unresolved measurements can also be used to investigate disk evolution. \cite{Ma+25} used optical forbidden line strengths, profiles, and ratios from highly irradiated, unresolved disks in $\unslant\sigma$-Orionis to demonstrate that irradiated disks appear more evolved, suggesting a shorter lifetime.

The disk fraction of a SFR is measured photometrically and is thus a more readily observable tracer of disk destruction. Many such studies have focused on nearby, lower-mass SFRs such as Taurus \citep[d$\sim$147~pc;][]{Ha+00, Zu+20}. The proximity to these SFRs allows for easier detection of lower-mass stars for a given sensitivity and angular resolution. However, these low-mass regions do not contain many OB stars and thus do not represent the star formation environment of the average Galactic star. While clusters massive enough to reliably form O stars ($\gtrsim1000$~M$_{\odot}$) are less common, the shape of the cluster mass function \citep[approximately $\frac{dN}{dM}\propto M^{-2}$;][]{LL03} means that a roughly equal number of stars form per dex of cluster mass and so a significant fraction of all stars form in massive clusters where external UV fields are high \citep{MS78}.

The disk lifetime is crucial for models of planet formation, as it determines the amount of time and material available for planet formation, effectively limiting the quantity and types of planets that can form \citep[e.g.,][]{Ni+19, Wi+22, Pa+23, Qi+23, Hu+24, HL25}. To better understand how external photoevaporation accelerates disk destruction and thus reduces the disk lifetime, we must measure the disk fraction in more massive and distant Galactic SFRs.

A few studies have explored the impact of external photoevaporation on the disk fraction in high-mass regions. For example, \cite{Pr1+11,Pr2+11}, collectively P11 hereafter, using JHKs photometry from the High Acuity Wide-field K-Band Imager \citep[HAWK-I;][]{HAWKI1, HAWKI2, HAWKI3, HAWKI4} on the Very Large Telescope (VLT), found that the highly-irradiated clusters of the Carina star-forming complex \citep[d$\sim$2.35~kpc;][]{GP22} have lower disk fractions than other clusters of similar age. Within individual high-mass SFRs, \cite{Ba+07}, \cite{Fa+12}, \cite{Gu+23}, and others have found a lower disk fraction close to OB stars/in the highest UV environments. However, others such as \cite{Ro+11}, \cite{Ri2+15}, \cite{Me+22}, and \cite{Da+24} have found, with similar data, no correlation between disk fraction and incident UV flux/distance to OB stars within individual SFRs. Much of this confusion comes from low completeness in distant, heavily extincted regions. Deeper studies of high-mass SFRs that account for observational biases are needed to resolve this discrepancy.

The above works use a combination of JHK ($\sim$$1.2-2.3~\unslant\mu$m) near-IR (NIR) and \textit{Spitzer} mid-IR (MIR; $\sim$$3.2-9.2~\unslant\mu$m) photometry to measure the disk fraction. Some YSOs can lose their hot inner disks before their cooler outer disks and thus require MIR observations to be observed \citep[e.g.,][]{va+23}. However, this is unlikely to occur in high-UV environments \citep[see e.g.,][]{Co+22, Ga+24}. Moreover, observations of distant, high-mass SFRs are challenging and require a higher sensitivity and resolution to match the completeness in nearby SFRs. This is difficult to achieve in the MIR, since bright, diffuse emission from H\,\textsc{ii} regions, combined with the $\sim$$2^{\prime\prime}$ resolution of \textit{Spitzer} limits the efficacy of existing MIR observations \citep[e.g.,][]{Po+09}. JHK observations do not possess these limitations, and thus are needed for the massive environments where external photoevaporation is most significant. Thus, we consider only JHK and JHKs excess fraction measurements for the disk fraction, using ``JHK(s) excess fraction,'' ``NIR excess fraction,'' and ``disk fraction'' interchangeably.

To our knowledge, only one team has performed a JHKs survey of a massive SFR complete down to the low-mass regime ($\lesssim0.5-1$~M$_{\odot}$). As mentioned, P11 surveyed the Carina star-forming complex, which contains many distinct clusters, the densest of which is Trumpler 14 (Tr~14). Tr~14 has a similar age to the ONC and Taurus ($1-2$~Myr), but the YSOs in Tr~14 experience significantly more UV radiation \citep[][]{Sm06, Be+23} from the $\sim$20 confirmed member O stars, especially from the double O2~I system HD~93129~A \citep{Be+23}. P11 measured a JHKs excess fraction of $9.7\pm0.8\%$ in Tr~14, significantly lower than the values they report for the ONC ($\sim$36\%, one O7V and one O9.5V star) or Taurus ($\sim$49\%, no OB stars). This difference between regions of similar age suggests that external UV radiation from the OB stars reduces the disk lifetime.

\begin{figure*}
    \centering
    \includegraphics[width=\textwidth]{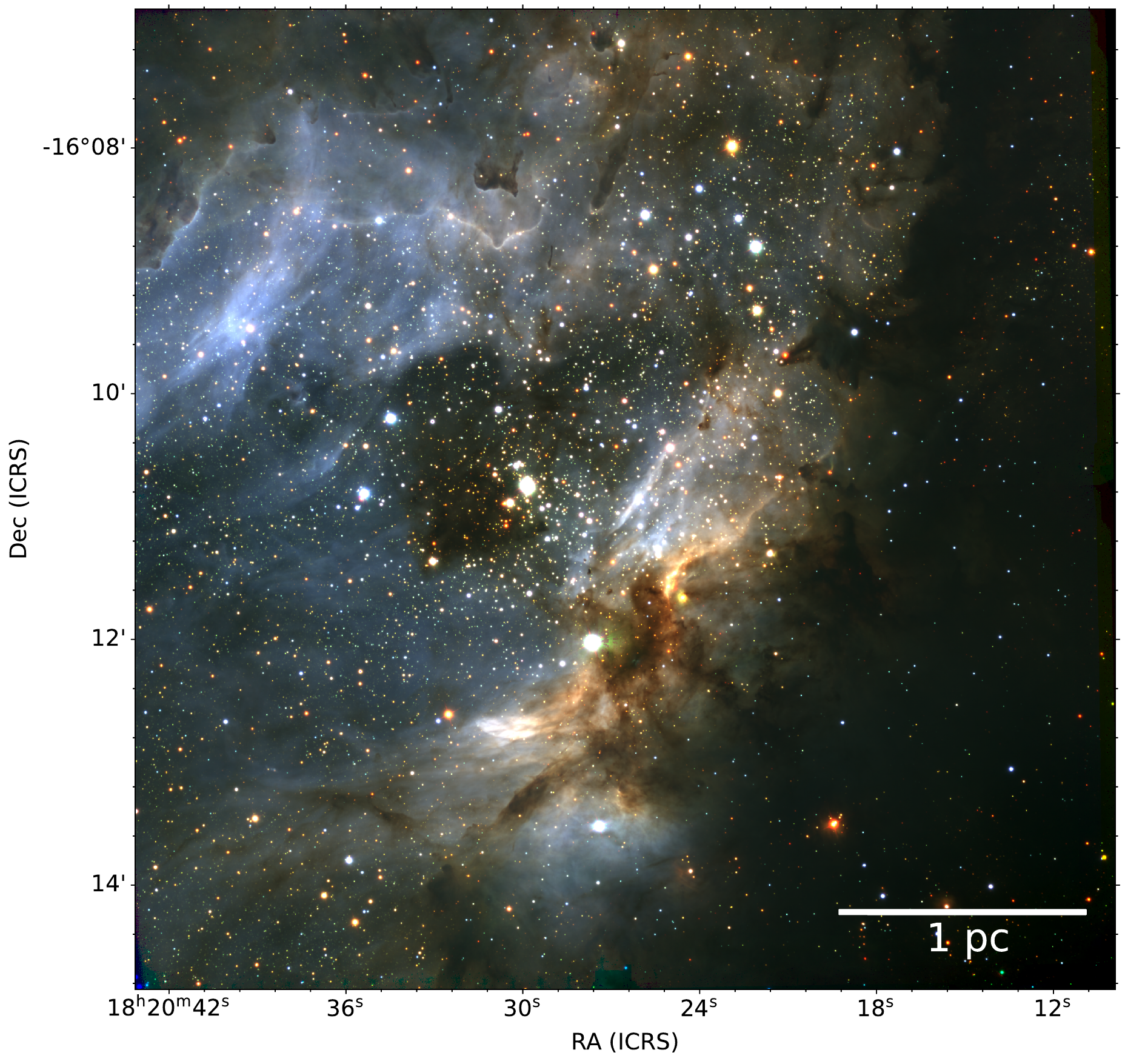}
    \caption{Near-infrared JHKs image of M17 from VLT/HAWK-I. More than 10,000 sources detected down to J-band magnitude of $\sim$$19.5-21$ with an angular resolution of $\sim\!0.3-0.5^{\prime\prime}$ (see Table \ref{tab:Observations}). The O4-O4 double binary CEN~1 located at $18^{\rm h}20^{\rm m}29.8^{\rm s}, ~-16^{\circ}10^{\prime}45^{\prime\prime}$ marks the center of the region, and note the IR-dark cloud directly to the east/southeast. Red: 2.15~$\unslant\mu$m, Green: 1.62~$\unslant\mu$m, Blue: 1.26~$\unslant\mu$m.}
    \label{fig:3-color}
\end{figure*}

In this work, we explore M17, pictured in Figure \ref{fig:3-color}. The M17 H\,\textsc{ii} region is located in Sagittarius at a distance of $\sim$1.7~kpc \citep{BP18, St+24}. At its center lies the young \citep[$\sim$1~Myr;][]{handbookM17}, massive ($\sim\!6000~{\rm M}_{\odot}$)\footnote{Estimated by Metropolis-Hastings sampling of the \cite{Ch03} IMF using the total stellar population estimate of 16000 by \cite{Ku+15}.} cluster NGC~6618, making M17 one of the nearest high-mass SFRs. M17 has eight confirmed member O stars, seven of which lie within a projected $\sim$1~pc radius of the center of the region, including the double O4-O4 spectroscopic binary CEN~1 \citep{Ch+80, St+24}. This produces a high degree of ionization relative to more nearby regions \citep{Gl98} and a variation in UV flux of over three orders of magnitude (log(G$_{0})=~\sim$$3.3-7.0$) across the central $\sim$4~pc~$\times$~4~pc region of the H\,\textsc{ii} region. The location of the O stars in the center of the region provides a simple UV environment with flux decreasing farther from the cluster center. The age, proximity, geometric simplicity, and intense UV environment make M17 an excellent region in which to investigate how the disk population is affected by different levels of external UV radiation.

Using HAWK-I, we measure the JHKs excess fraction in an $\sim$$8^{\prime}\times8^{\prime}$ ($\sim$4~pc~$\times$~4~pc) region towards the center of M17. Previous NIR observations of M17 in the Two-Micron All Sky Survey \citep[2MASS,][]{2MASS} and UKIRT Infrared Deep Sky Survey \citep[UKIDSS,][]{UKIDSS} lacked the physical resolution and sensitivity necessary to detect many low-mass sources. The survey presented here is $\sim$1.5~mag deeper and $\sim$$2-3$~times higher-resolution than UKIDSS and provides new NIR photometry to 10,339 sources. Of these sources, 1016 are present in the Massive Young Star-Forming Complex Study in Infrared and X-ray \citep[MYStIX,][see Section \ref{sec:MYStIX}]{Fe+13, Br+13} X-ray catalog, including 117~sources (12\%) without any previous NIR photometry before this work.

\section{Data} \label{sec:data}
We present new VLT/HAWK-I JHKs images ($1.18-2.31~\unslant\mu$m, ESO program 111.252G, P.I. M.~Reiter). We also use the MYStIX Probable Complex Member (MPCM) catalog to select YSOs and calibrate our photometry. 

\subsection{HAWK-I}
HAWK-I is a wide-field NIR ($0.85-2.5$~$\unslant\mu$m) imager on the UT4 telescope of the VLT. It has a field of view (FOV) of $7.5^{\prime} \times 7.5^{\prime}$ with a 15$^{\prime\prime}$ gap between each of its four detectors. For this work, we observed in the J ($\lambda_{\rm cen}=1.258~\unslant\mu$m), H ($\lambda_{\rm cen}=1.620~\unslant\mu$m), and Ks ($\lambda_{\rm cen}=2.146~\unslant\mu$m) broadband filters with 36, 12, and 8 exposures, respectively, each with 3 or 6 integrations of 20 or 10 seconds each, respectively, dithered to cover the detector gap and provide a larger imaged area ($\sim$$8^{\prime}\times8^{\prime}$, $\sim$4~pc~$\times$~4~pc). The Ground Layer Adaptive Optic mode \citep[][]{HAWKI5, HAWKI6} provided an angular resolution of $\sim\!0.3-0.5^{\prime\prime}$ (see Table \ref{tab:Observations}).

\begin{deluxetable*}{ccccccc}
    \tablecaption{Summary of HAWK-I Observations\label{tab:Observations}}
    \tablehead{
    \colhead{Date} & \colhead{Filter} & \colhead{NIR FWHM ($^{\prime\prime}$)} & \colhead{\# of Frames} & \colhead{NDIT} & \colhead{DIT (s)} & \colhead{Eff. Exp. Time (s)}
    }
    \startdata
    May-21 & Ks & 0.36 & 8  & 6 & 10 & 480 \\
    Jun-28 & J  & 0.32 & 12 & 3 & 20 & 960 \\
    Jun-29 & J  & 0.34 & 12 & 3 & 20 & 960 \\
    Jul-09 & H  & 0.34 & 12 & 6 & 10 & 960 \\
    Jul-26 & J  & 0.48 & 12 & 3 & 20 & 960 \\
    \enddata
    \tablecomments{All observations from 2023. NIR FWHM is the average measured FWHM using \texttt{imexamine} on various isolated sources in each final mosaic. NDIT refers to the number of sub-integrations per exposure, and DIT is the integration time per sub-integration.
    }
\end{deluxetable*}

\subsection{MYStIX} \label{sec:MYStIX}
YSOs are very bright X-ray sources. Violent magnetic reconnection from convection excites coronal plasma and produces significant X-ray radiation\footnote{While A and late B type stars do not produce significant X-rays \citep[e.g.,][]{Ev+11}, in this work we predominantly consider sources $<1.8$~M$_{\odot}$, of approximate type F and later.} \citep{PF02}. These flares indicate youth, suggesting that these sources are cluster members. This is a strong selection criterion, since field objects tend to be older and produce significantly less X-ray emission \citep{Fe+07}\footnote{Although main-sequence M-type stars also exhibit magnetic activity and produce X-ray emission, the typical X-ray fluxes are much weaker than those observed in pre-main-sequence stars of similar spectral types.}. Therefore, we use the MYStIX MPCM catalog to select the YSOs in M17. MYStIX is a survey of 20 OB-dominated young clusters and their surrounding environments in the nearest 4~kpc, including the entire HAWK-I FOV observed here, using X-ray observations from the \textit{Chandra X-ray Observatory} and infrared observations from the \textit{Spitzer Space Telescope}, United Kingdom InfraRed Telescope (UKIRT), and 2MASS combined with spectroscopically confirmed OB stars from the literature.

\subsection{\textit{Spitzer}}
MIR emission traces cooler material farther out in the disk, so is generally preferred when measuring the impact of external photoevaporation. However, the JHK data presented here have $\gtrsim4$ times higher resolution and much greater sensitivity than the available MIR data of M17 from \textit{Spitzer} \citep[][]{Po+09}. Specifically, we find only 50 sources with high-quality JHKs and \textit{Spitzer} 3.6~$\unslant\mu$m photometry in the HAWK-I FOV that match with MYStIX sources, compared to the 932 high-quality JHKs sources presented in Section \ref{sec:isochrones}. Due to this incompleteness, we do not consider \textit{Spitzer} sources in M17 in this work.

\subsection{Data Reduction}
\subsubsection{EsoRex and the HAWK-I Pipeline}
We performed the initial data reduction using the ESO VLT data-flow system through EsoRex\footnote{\url{https://www.eso.org/sci/software/cpl/esorex.html}} \citep{EsoRex}. The science and sky frames were dark-corrected and flat-fielded, and sky emission was removed using the \texttt{pawsky\_mask} algorithm (see the HAWK-I pipeline user manual for more information). The pipeline calibrated the astrometry against 2MASS sources in each image and produced calibrated images corresponding to each raw image.

\subsubsection{Mosaicking}
The calibrated images then had to be combined into the final mosaics. However, each frame had an additive flux offset in the pixel values at the same sky location between frames that first had to be corrected. Because of the extended and variable nebulosity in M17, EsoRex could not correct these offsets. To properly take the large-scale nebulosity across the field into account before combining the images, we devised our own mosaicking method using modified code from the \texttt{reproject}\footnote{\url{https://reproject.readthedocs.io}} Python package.

First, using a modified version of the \texttt{find\_optimal\_celestial\_wcs} method, we determined the ideal world coordinate system (WCS, the object that contains information for geometric transformations between coordinate systems) and shape for the output array of the final mosaic based on the footprint of the input frames. We then used a modified version of the \texttt{reproject\_and\_coadd} mosaicking method to retrieve the relative corrections for the pixel-value flux offsets between each frame. We subtracted these corrections from the frames, which forced them all to the same background level. Then, we used \texttt{reproject\_exact} to reproject the frames while conserving the correct amount of flux in each pixel, accounting for the distortion of taking flat images of the spherical sky. Using this method, we constructed a \texttt{numpy} array where each element was one reprojected frame aligned to the same WCS. We mean-combined the reprojected images after sigma-clip rejection with a sigma of 10.

\section{Methods} \label{sec:methods}
\subsection{Source Detection \& Photometry}
For each filter, we identified point sources using the \texttt{DAOFIND} implementation in the \texttt{photutils} Python package \citep{dao,photutils}. We searched for sources at least 3$\unslant\sigma$ above the background, determined in a dark area of the J-band image. To identify only point sources and remove noise spikes and extended nebulosity, we imposed \texttt{sharpness} and \texttt{roundness} limits based on visual inspection. This method was performed conservatively to avoid removing any true sources while simultaneously significantly reducing false matches during catalog matching (see below).

We performed aperture photometry with a circular radius of 2.5~pixels (corresponding to $0.27^{\prime\prime}$), and an inner/outer annular radius of 15/20~pixels for background estimation. These radii were determined by examining the curve of growth of the magnitude and magnitude error of an isolated test star. 2.5~pixels provided a low fractional error on the magnitude calculation while being simultaneously approximately half the largest observed full width at half maximum (FWHM). The total error on the measurements was calculated using the \texttt{calc\_total\_error} method from \texttt{photutils}, which combines background error with the Poisson noise from the source, where the background error was estimated to be the standard deviation of sigma-clipped pixel values in the annuli noted above. We sigma-clipped the background values in the annuli with a sigma of 3 to avoid contamination from other sources in crowded regions.

\begin{figure}
    \centering
    \includegraphics[width=1\linewidth]{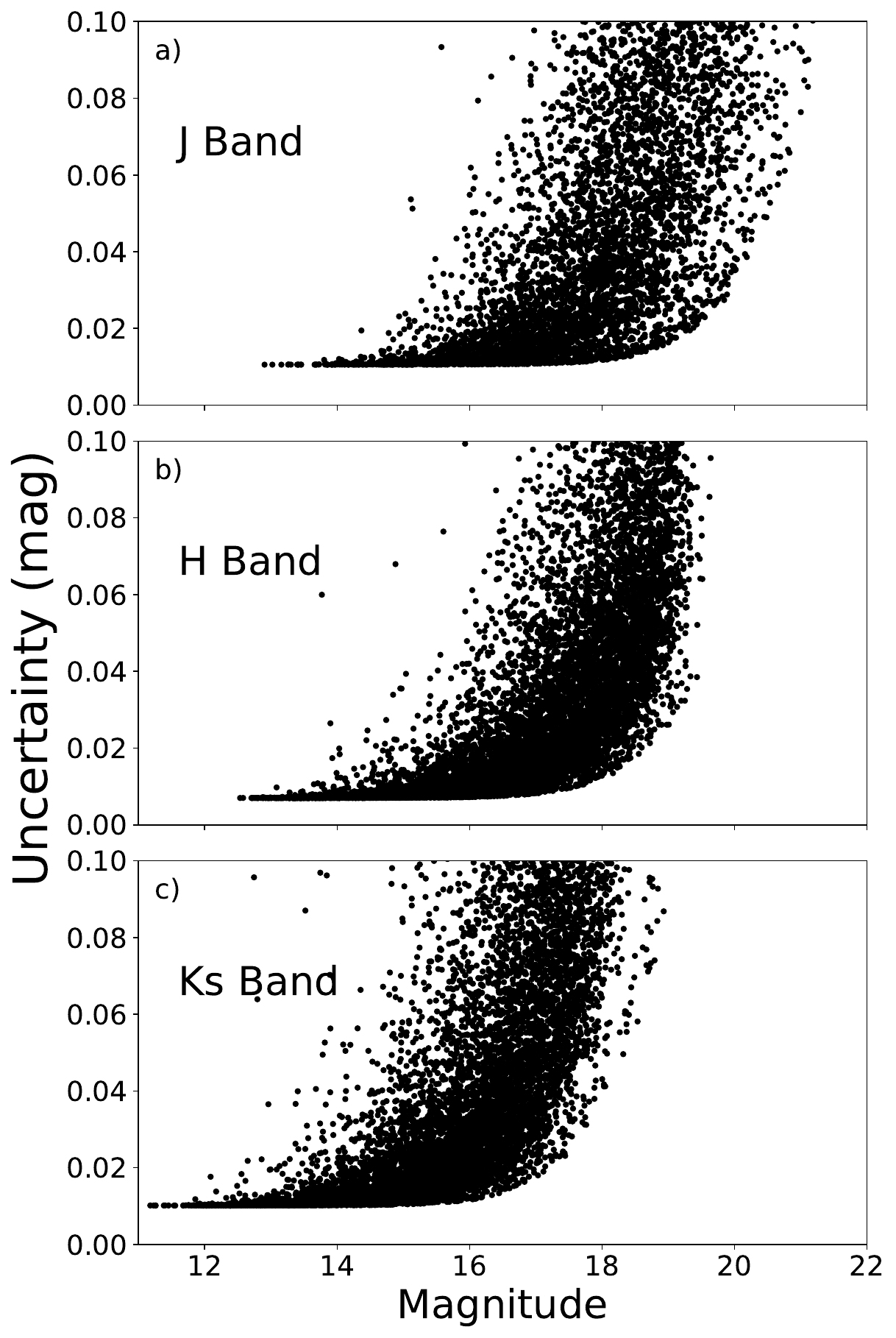}
    \caption{Magnitude vs. uncertainty on the magnitude in a) J, b) H, and c) Ks bands for all sources detected simultaneously in J, H, and Ks. The minimum uncertainty of $\sim\!0.01$~mag comes from uncertainty on the ZP. The larger scatter in magnitude in the J-band is likely due to the increased nebulosity at shorter wavelengths. Survey depth is $\sim$$19.5-21$~magnitudes in J-band, depending on position.}
    \label{fig:mag_v_mag_err}
\end{figure}

We catalog matched the J, H, and Ks bands using \texttt{match\_to\_catalog\_sky} from \texttt{astropy} with a match tolerance of $0.15^{\prime\prime}$ (approximately half the smallest observed FWHM) to identify sources detected in all three bands. This resulted in a final catalog of 10,339~sources in common between the three bands. The increased scatter in magnitude for J-band detections seen in Figure \ref{fig:mag_v_mag_err} is likely due to the increased nebulosity at shorter wavelengths.

We use the calibrated UKIDSS JHK photometry from the MYStIX MPCM catalog to calibrate our photometry. To calculate the zeropoints (ZPs) for each band, we first selected only MYStIX sources with existing UKIDSS JHK photometry with photometric uncertainties $<0.1$~mag\footnote{All uncertanties propagated with the \texttt{uncertainties} Python package unless stated otherwise (https://pythonhosted.org/uncertainties/).}. Then, we catalog matched with our observations using a 0.15$^{\prime\prime}$ selection tolerance, revealing 522 high-quality MYStIX sources in the FOV. The low tolerance for matches compared to the resolution of \textit{Chandra} was to prioritize matching isolated sources in our catalog with single MYStIX sources rather than matching as many sources as possible. Next, we subtracted the instrumental magnitudes from the corresponding MYStIX magnitudes. This led to aperture-corrected median ZPs of $24.44\pm0.01$, $24.95\pm0.01$, $23.78\pm0.01$ mag for the J, H, and Ks bands, respectively. These ZPs were added back to the instrumental magnitudes to achieve the final magnitudes for each source.

\subsection{Member Selection}
\label{sec:member selection}
Because M17 lies in the Galactic plane, there is significant fore/background contamination along the line of sight. We use the MYStIX MPCM catalog, which consists of spectroscopically confirmed OB stars and X-ray-active YSOs, to distinguish cluster members from field sources. \cite{Br+13} describe how they use probabilistic identifiers such as UKIDSS J-band magnitude; median X-ray energy, variability, and hardness ratio; distance from other X-ray sources; SED modeling; etc., to identify contaminants. They found 703 of the 2999 X-ray sources towards M17 to be contaminants, which agrees with their simulations. Assuming contaminants are randomly distributed across the field \citep[unlike cluster members, see Figure 4 of][]{Br+13}, and taking into account the smaller FOV in this work, we can expect $\sim$$50-60$~contaminants ($\sim$$5-6\%$) in our FOV. The MYStiX MPCM catalog does not contain these 703 contaminants, so the estimate here should be taken as a very conservative upper limit for the number of X-ray contaminants in our final catalog; the true number should be much smaller.

\begin{figure}
    \centering
    \includegraphics[width=0.4\textwidth]{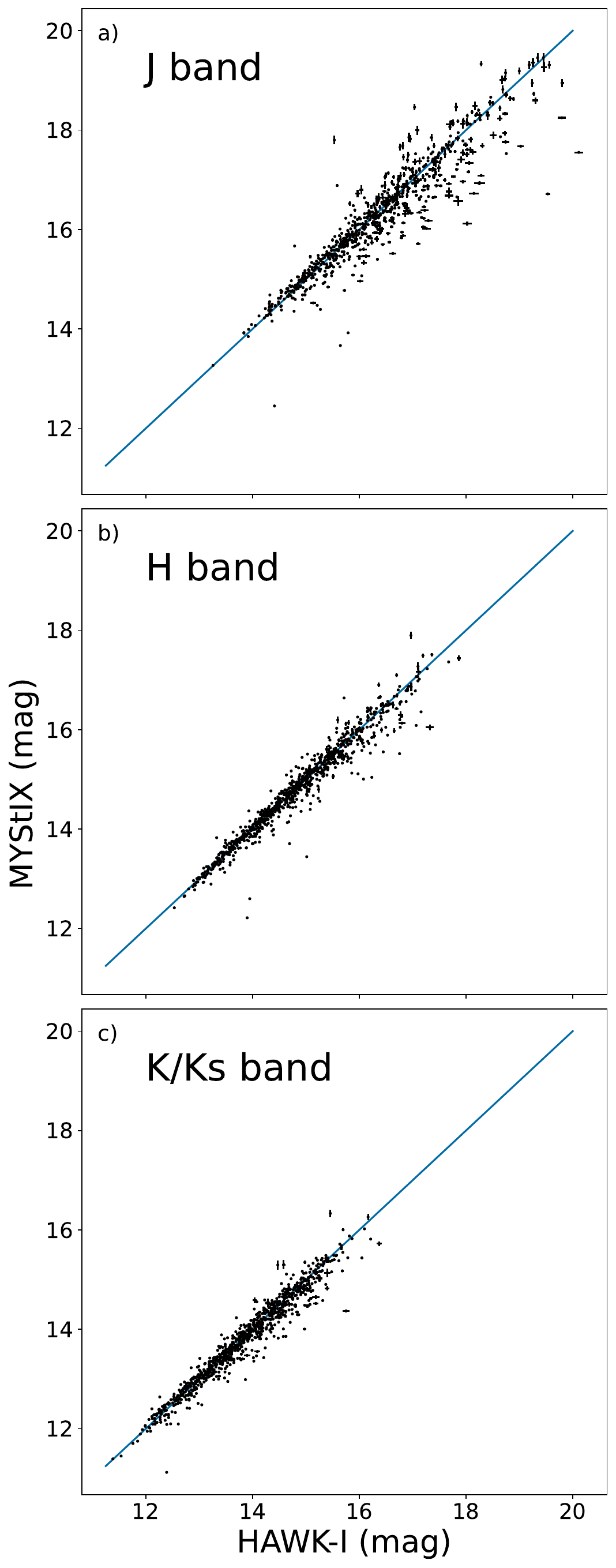}
    \caption{Photometry comparison between HAWK-I and MYStIX/UKIDSS in the a) J b) H c) K (MYStIX)/Ks (HAWK-I) band. The blue line represents a 1:1 correlation. While there is some scatter, especially in J-band, there is strong agreement between the two sets of photometry. Much of the differences can be attributed to the intrinsic variability of YSOs.}
    \label{fig:phot_comparison}
\end{figure}

The limiting magnitude and resolution in previous NIR studies\footnote{$\sim$$18-19.5$~mag J-band limiting magnitude dependent on position, $\lesssim1^{\prime\prime}$ ($\lesssim2000$~AU) resolution from UKIRT \citep[][]{Ki+13, UKIDSS}.} left many MYStIX sources undetected in the NIR. We complement the catalog with the more sensitive and higher-resolution HAWK-I observations\footnote{$\sim$$19.5-21$~mag J-band limiting magnitude dependent on position, $\sim\!0.3-0.5^{\prime\prime}$ ($\sim\!600-1000$~AU) resolution from this work.} presented here. Matching with the full MYStIX catalog with a 0.5$^{\prime\prime}$ selection tolerance (to match the best angular resolution of \textit{Chandra}) and picking the closest matches revealed 1016 cluster members in the FOV, including 117~sources without previous MYStIX NIR photometry. For a comparison between the 851 MYStIX sources detected with both UKIDSS and HAWK-I with photometric uncertainties $<0.1$~mag, see Figure \ref{fig:phot_comparison}. \cite{Fi+23} assert that YSOs routinely vary by $\lesssim1-2$~mag in the optical/NIR, which accounts for 97\% and 100\% of the variation in the J- and  H-/K(s)-bands, respectively. The remaining J-band scatter is likely due to the stronger nebulosity and YSO variability at shorter wavelengths.

In the full HAWK-I catalog of 10,339 sources, 1184 sources (11\%) have at least one neighbor within $1^{\prime\prime}$ and 443 (4\%) within $0.8^{\prime\prime}$. A significant fraction of these neighbors appear as single sources in UKIDSS, given the stated $\lesssim1^{\prime\prime}$ resolution. In the catalog of MYStIX-selected sources with photometric uncertainties $<0.1$~mag in J, H, and Ks, 18 sources (2\%) have a neighbor within $1^{\prime\prime}$ and 12 (1\%) within $0.8^{\prime\prime}$. The lower percentage comes from the fact that the MYStIX X-ray and UKIDSS data have a comparable resolution.

\begin{figure*}
    \centering
    \includegraphics[width=0.8\linewidth]{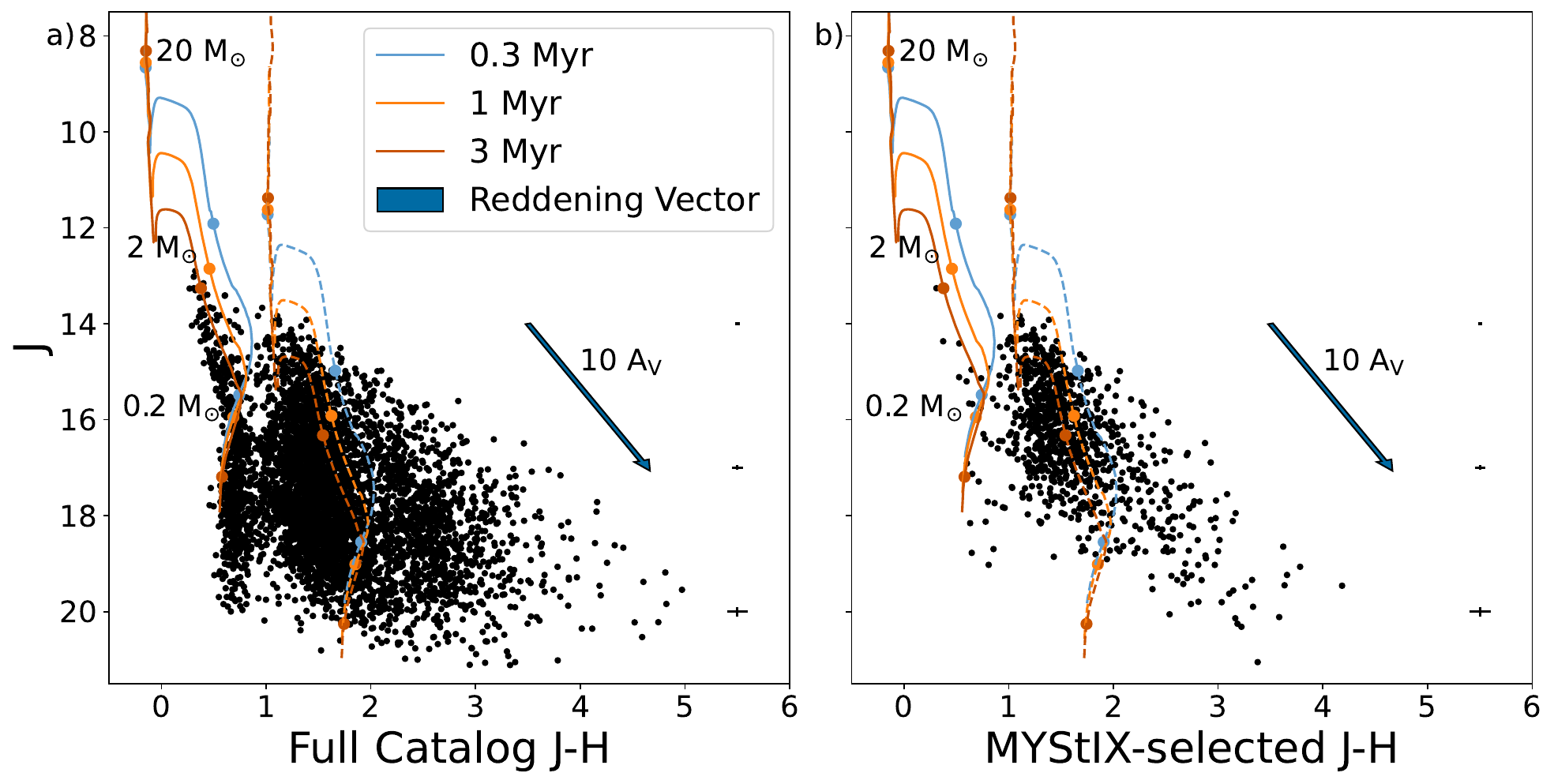}
    \caption{$(J-H)$ vs. $J$ color-magnitude diagrams for sources with photometric uncertainties $<0.1$~mag in a) the full catalog (4727 sources) and b) the catalog of sources with counterparts in MYStIX (932 sources). Typical uncertainties shown to the right. Overlaid are PARSEC 1.2S 0.3, 1, and 3~Myr isochrones without extinction (solid lines) and with $A_{V}=10$~mag (dashed lines). Isochrone masses of 0.2, 2, and 20~M$_{\odot}$ are marked with filled circles. The arrow represents an $R_{V}=4$, $A_{V}=10$~mag \cite{In+05} reddening vector (see Section \ref{sec:dereddening}).}
    \label{fig:isochrones}
\end{figure*}

\begin{figure*}
    \centering
    \includegraphics[width=0.8\linewidth]{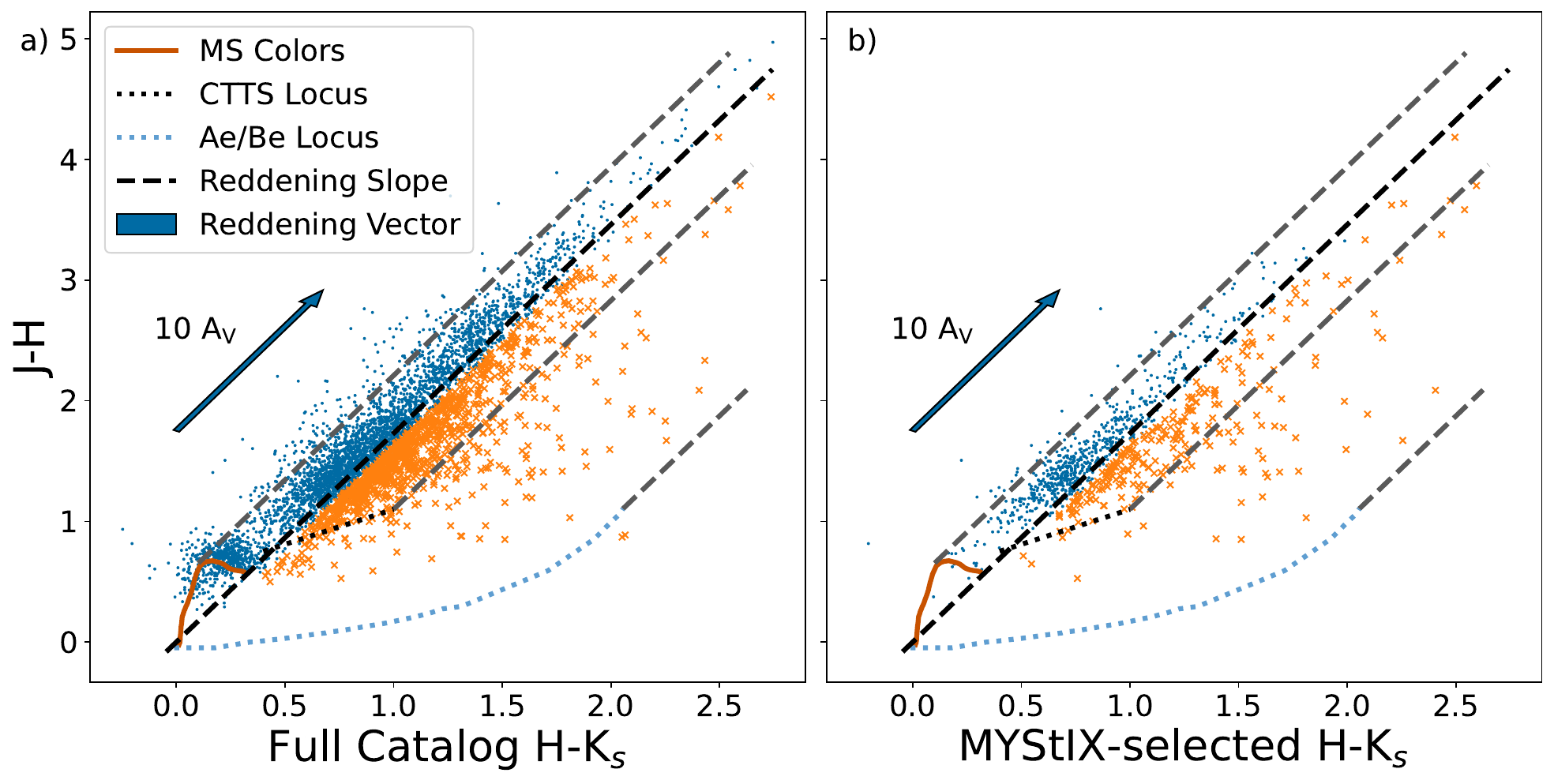}
    \caption{$(J-H)$ vs. $(H-K_{\rm s})$ color-color diagrams for sources as in Figure \ref{fig:isochrones}. The isolated population to the lower left in (a) consists mainly of foreground sources. The dark orange curve is a 100~Myr PARSEC isochrone $\leq2$~M$_{\odot}$ representing main sequence (MS) colors. The black dotted line is the classical T Tauri star (CTTS) locus of \cite{Me+97}. The light blue dotted line is the Herbig Ae/Be star locus of \cite{LA92}. The black dashed line is the \cite{In+05} reddening slope of 1.73, with parallel gray dashed lines extending from the MS colors and the CTTS and Herbig Ae/Be loci. The arrow is an $R_{V}=4$, $A_{V}=10$~mag reddening vector along the direction of the reddening slope. a) $22.9\pm0.8\%$ and b) $28\pm2\%$ of sources lie at least 1$\unslant\sigma_{\rm phot}$ and 0.05~mag below and to the right of the reddening slope (orange Xs) and have JHKs excess (see Section \ref{sec:IR excess}.)}
    \label{fig:populations}
\end{figure*}

\subsection{CMDs and CCDs} \label{sec:CMD/CCD}
\subsubsection{Isochrones and Reddening} \label{sec:isochrones}
Figures \ref{fig:isochrones} \& \ref{fig:populations} show the color-magnitude and color-color diagrams (CMDs and CCDs), respectively, for the full and MYStIX-selected populations in the observed field. With these, we identified young sources with JHKs excess in M17 and compared with 0.3, 1, and 3~Myr isochrones from the PARSEC evolutionary model\footnote{http://stev.oapd.inaf.it/cgi-bin/cmd} \citep[ver. 1.2S;][]{Ma+17}. In both figures, we have plotted only those sources with photometric uncertainties $<0.1$~mag in J, H, and Ks. This leaves 4727 out of 10,339 sources for the full low-uncertainty catalog, and 932 out of 1016 for the MYStIX-selected low-uncertainty catalog. In these figures, it is clear that there are two distinct populations visible on the CCDs and CMDs. We believe that the smaller group, to the left in both figures, represents the population of older foreground stars in the FOV, since this group mostly disappears when matching with the MYStIX sources, is consistent with the unreddened main sequence colors on the CCD, and experiences low levels of extinction. This conclusion is also supported by comparison with the Besançon stellar population synthesis model of the Galaxy \citep{Ro+03}. The extended population consists mostly of cluster members and background objects extincted to varying degrees due to the nebulosity of the region.

\subsubsection{Dereddening}
\label{sec:dereddening}
We use the 1~Myr isochrone to estimate the mass of and extinction towards each object in M17 by dereddening. The sources are dereddened to the isochrone along the reddening vector of P11 and \cite{Po+11}, which describes how a source moves in color-magnitude or color-color space under a given amount of reddening (see Figure \ref{fig:isochrones}). This reddening vector uses the \cite{In+05} extinction law (the functional form for how much light is extincted at a given wavelength), with its length scaled to represent $R_{V}$, the ratio of total to selective extinction, equal to 4. This matches well with $R_{V}=3.9$ for M17 given by \cite{handbookM17}. By tracing the source to the isochrone along the reddening vector, or, equivalently, by tracing from the isochrone along the reddening vector to the source, the corresponding dereddened position of the source on the isochrone can be determined. We note that for the chosen law, changing $R_{V}$ impacts only the magnitude of the reddening vector; not its direction.

Since the isochrone consists of a series of discrete points, the mass of each source must be interpolated from those discrete values. The estimated mass is the distance-weighted average of the two points nearest to the YSO on the reddened isochrone. To estimate uncertainties on the derived mass, we employed the random sampling method of \cite{Ro+26} to generate an empirical uncertainty relationship:

\begin{equation}
    \sigma_M=0.4(\frac{\sigma_{(J-H)}}{\sqrt{0.02}})^{\frac{3}{4}}M,
\end{equation}
where $M$ is the mass and $\sigma_{(J-H)}$ is the uncertainty on the $(J-H)$ color.\footnote{Because the isochrone is nearly vertical, the uncertainty on the J-band magnitude has little effect on the uncertainty in the mass or extinction.}

\begin{figure*}
    \centering
    \includegraphics[width=0.8\textwidth]{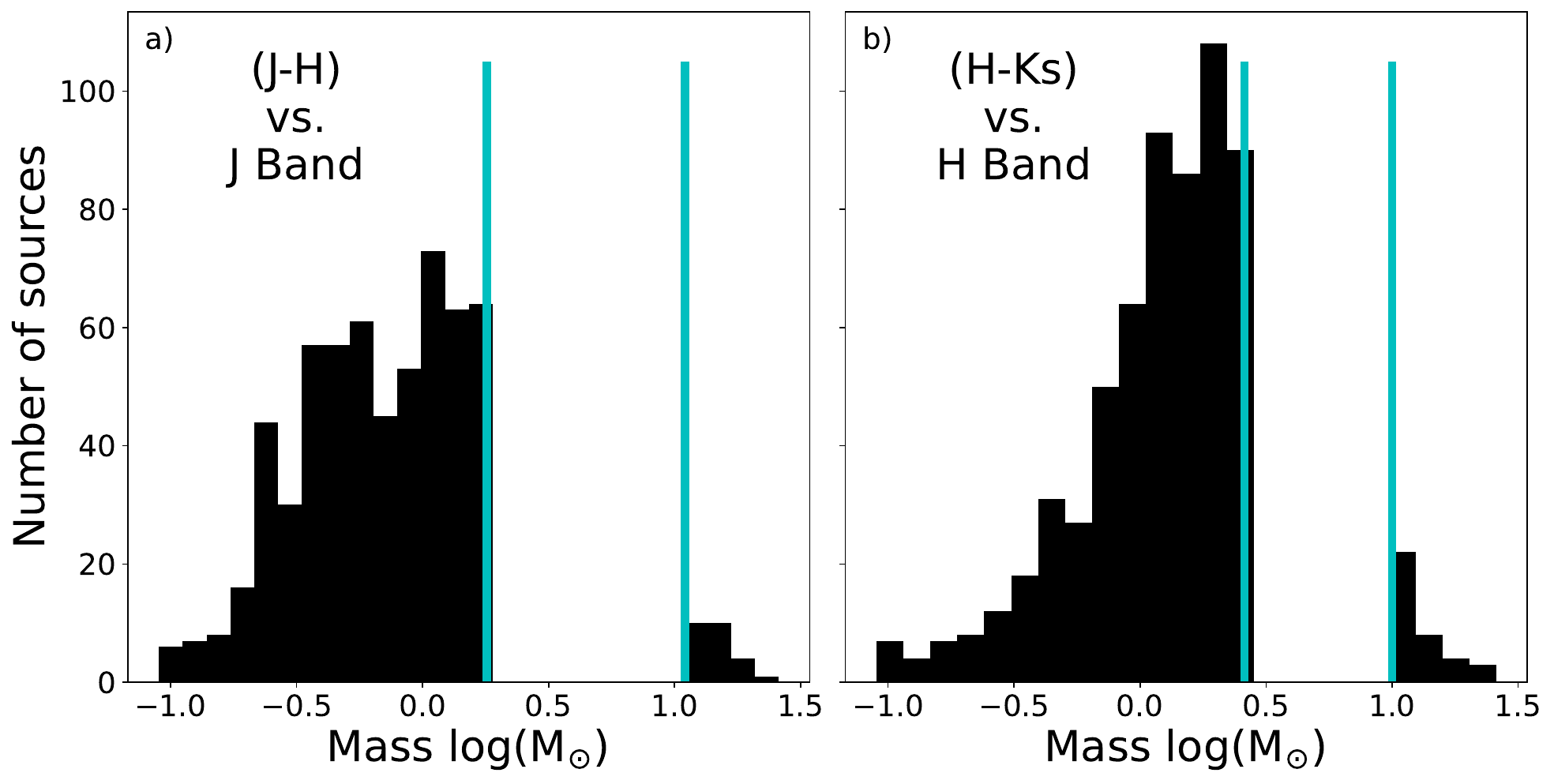}
    \caption{Histogram of mass estimates for MYStIX-selected sources in M17 using a 1~Myr PARSEC isochrone from the a) $(J-H)$ vs. $J$ and b) $(H-K_s)$ vs. $H$ CMD. Because the isochrone has a kink on both CMDs, there is a region in both figures where the mass estimates are degenerate, marked by cyan lines. For sources in this region, we report the mass corresponding to the smallest $A_{V}$ in Table \ref{tab:Data}, though they are excluded here, for clarity.}
    \label{fig:Masses}
\end{figure*}

The isochrones are not monotonic in the mass range where stars transition from pre-main sequence to main sequence objects (see Figure \ref{fig:isochrones}). For this reason, dereddening a source in a specific mass range can lead to multiple locations on the isochrone. For the $(J-H)$ versus $J$ isochrone, this ``region of degeneracy'' exists between 1.8 and 11~M$_\odot$, while for the $(H-K_{\rm s})$ versus $H$ isochrone, it exists between 2.6 and 10~M$_\odot$. Figure \ref{fig:Masses} shows the derived masses, and we exclude sources without a unique mass determination from the figures and any mass-dependent analysis in this work for clarity. In Table \ref{tab:Data}, we report the mass corresponding to the intersection on the isochrone with the smallest $A_{V}$.

\begin{figure*}
    \centering
    \includegraphics[width=0.8\textwidth]{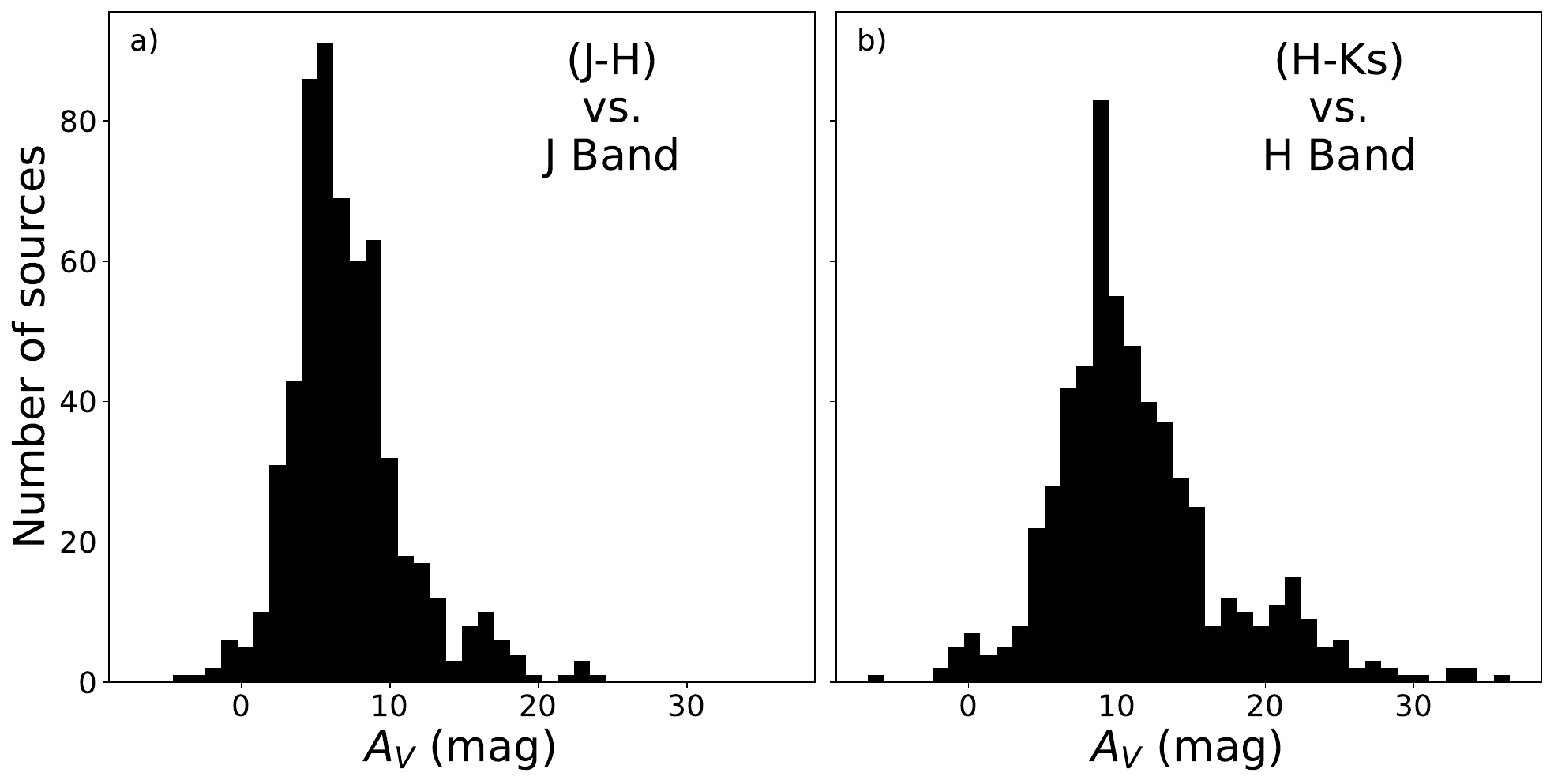}
    \caption{Histograms of extinction estimates for MYStIX-selected sources $<1.8~{\rm M}_{\odot}$ in M17 using a 1 Myr PARSEC isochrone calculated from the a) $(J-H)$ vs. $J$ b) $(H-K_s)$ vs. $H$ CMD}.
    \label{fig:Extinction}
\end{figure*}

The distance traveled relative to the length of the reddening vector needed to intersect the isochrone determines the extinction for each source. See Figure \ref{fig:Extinction} for the derived visual extinctions. The uncertainty on the extinction comes from an empirical relationship developed in the same manner as the one for the mass:
\begin{equation}
    \sigma_{A_V}=2(\frac{\sigma_{(J-H)}}{\sqrt{0.02}})~\rm{mag},
\end{equation}
where $\sigma_{(J-H)}$ is as above.

\subsection{UV Field}
To calculate the UV field in M17 as a function of spatial position, we determined the FUV \citep[912 to 2400~\AA;][]{Ha68} luminosity of all of the confirmed OB stars in M17 using the catalog of \cite{St+24}, who give membership and spectral-type estimates for each O and B star in the region. We used the conversions given by \cite{Ma+05} to convert the spectral types to temperatures and radii. Then, we integrated a blackbody model over the FUV range to get the emitted power per area and integrated over the surface area to get FUV luminosity per star.

In this work, we neglect EUV radiation because FUV radiation controls the mass-loss rate for most of the disk lifetime \citep[e.g., Figure 5 of][]{Ha+23, St+99, Ri+00, Ad+04, Wi+18}. To map out the UV field across M17, we added up the FUV contributions at each point from all the O stars, assuming the inverse square law and the projected 2D distance, converting to units of G$_{0}$ \citep[where 1~G$_{0}$ is equal to the FUV field in the solar neighborhood;][]{Ha68}. We used the same method to calculate the incident FUV flux at the position of each source, finding UV field strength of log(G$_0)\approx3.3-7.0$ in the imaged region.

We must clarify that, first, using 2D projected distances for a 3D distribution provides an upper limit for the FUV field. \cite{An+25} showed that 2D projection tends to overestimate FUV flux incident on individual YSOs by a factor of $\sim$2 at a distance of $\sim\!0.5$~pc from the cluster center, with better agreement at larger distances, though this depends on the geometry of the region. Second, this method assumes no intervening extinction between the OB stars and the YSOs, which is unlikely given the complex environment of M17. However, using the extinction maps of \cite{La+22}, \cite{An+25} tested the effect of shielding by intervening extinction and found a negligible ($\lesssim1\%$) total FUV flux decrease in the highest-UV regions. Third, we only include the confirmed member O stars and none of the candidates presented in \cite{St+24}, so our calculated UV field represents a lower limit of this 2D projection method.

\begin{figure}
    \centering
    \includegraphics[width=0.423\textwidth]{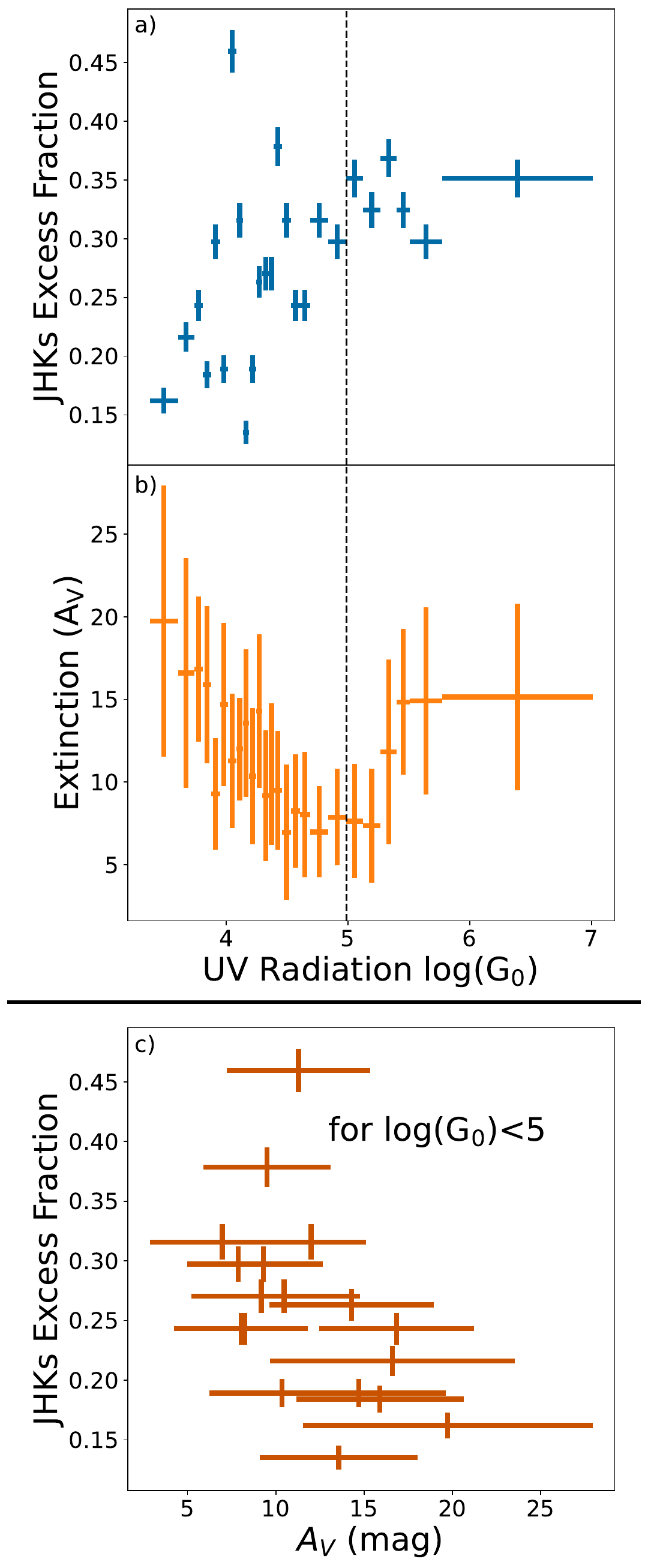}
    \caption{a) JHKs excess fraction and b) median extinction per bin vs. incident UV radiation. Each of the 25 bins contains $\sim$37 MYStIX-selected sources. Horizontal error bars: bin width; vertical error bars: a) calculated from $\sqrt{N}$, b) median absolute deviation. We observed a similar positive NIR excess trend for non-MYStIX-selected sources. The region to the right of the dashed line represents the center of M17 and the location of the IR-dark cloud (see Figure \ref{fig:X-ray Excess and UV}). c) JHKs excess fraction vs. median extinction for bins with $\rm log(G_{0})<5$, with the same errorbars as in (a), showing a general negative trend (see Section \ref{sec:DF v. UV}).}
    \label{fig:DF vs. UV}
\end{figure}

To determine the disk fraction as a function of UV radiation strength, we ordered the sources by the amount of UV radiation they receive, separated them into bins of equal number, and finally measured the JHKs excess fraction in each bin. Figure \ref{fig:DF vs. UV} shows that there is a positive correlation between disk fraction and incident UV radiation, which is naively unexpected, since UV radiation accelerates disk destruction (see Sections \ref{sec:DF v. UV} \& \ref{sec:DF v. UV Discussion} for further discussion of this phenomenon).

\subsection{Data Table}
A sample version of the final data table is given in Table \ref{tab:Data}. The full machine-readable table is available via the online journal.

\begin{deluxetable*}{cccccccccccccccc}
    \digitalasset
    \tablecaption{Photometry and Other Properties of Sources in M17\label{tab:Data}}
    \tablehead{
    \colhead{ID} & \colhead{MPCM\tablenotemark{a}} & \colhead{RA} & \colhead{Dec} & \colhead{m$_{\rm J}$} & \colhead{e\_m$_{\rm J}$} & \colhead{m$_{\rm H}$} & \colhead{e\_m$_{\rm H}$} & \colhead{m$_{\rm Ks}$} & \colhead{e\_m$_{\rm Ks}$} & \colhead{Excess\tablenotemark{b}} & \colhead{Mass\tablenotemark{c}} & \colhead{e\_Mass\tablenotemark{d}} & \colhead{$A_V$} & \colhead{e\_$A_V$\tablenotemark{d}} & \colhead{UV$_{\rm lum}$\tablenotemark{e}}
    }
    \startdata
     &  & $\mathrm{{}^{\circ}}$ & $\mathrm{{}^{\circ}}$ & $\mathrm{mag}$ & $\mathrm{mag}$ & $\mathrm{mag}$ & $\mathrm{mag}$ & $\mathrm{mag}$ & $\mathrm{mag}$ &  & $\mathrm{M_{\odot}}$ & $\mathrm{M_{\odot}}$ & $\mathrm{mag}$ & $\mathrm{mag}$ & $\mathrm{mW~m^{-2}}$ \\
     \hline
        ... & ... & ... & ... & ... & ... & ... & ... & ... & ... & ... & ... & ... & ... & ... & ... \\
        10220 & --- & 275.10251 & -16.11778 & 15.74 & 0.01 & 13.314 & 0.007 & 11.99 & 0.01 & N & 16.0 & 1.0 & 22.0 & 0.2 & 8.96 \\
        10221 & --- & 275.09218 & -16.11779 & 17.28 & 0.02 & 15.64 & 0.01 & 14.59 & 0.02 & Y & 0.47 & 0.05 & 7.0 & 0.4 & 7.87 \\
        10222 & --- & 275.1209 & -16.11774 & 17.66 & 0.04 & 16.88 & 0.02 & 16.63 & 0.03 & N & 0.11 & 0.02 & 1.9 & 0.6 & 10.04 \\
        10223 & 182019.23-160703.9 & 275.08017 & -16.11777 & 20.7 & 0.2 & 17.36 & 0.02 & 15.23 & 0.02 & Y & 1.6 & 0.8 & 24.0 & 3.0 & 6.48 \\
        10224 & --- & 275.13287 & -16.11774 & 21.2 & 0.6 & 15.363 & 0.007 & 12.72 & 0.01 & N & 100 & --- & 51.0 & 8.0 & 10.11 \\
        10225 & --- & 275.12448 & -16.11774 & 18.95 & 0.08 & 18.28 & 0.05 & 17.9 & 0.2 & N & -1 & --- & 4.0 & 1.0 & 10.12 \\
        10226 & 182019.97-160703.7 & 275.08327 & -16.11774 & 20.2 & 0.1 & 17.13 & 0.01 & 15.34 & 0.01 & N & 1.3 & 0.5 & 22.0 & 2.0 & 6.82 \\
        10227 & --- & 275.11198 & -16.11768 & 17.8 & 0.1 & 15.84 & 0.03 & 14.61 & 0.05 & N & 0.6 & 0.3 & 10.0 & 2.0 & 9.64 \\
        10228 & --- & 275.091 & -16.11768 & 20.4 & 0.2 & 18.2 & 0.03 & 17.12 & 0.06 & N & 0.17 & 0.08 & 14.0 & 2.0 & 7.71 \\
        10229 & --- & 275.07876 & -16.11766 & 20.6 & 0.4 & 17.59 & 0.04 & 15.85 & 0.04 & N & 0.6 & 0.6 & 19.0 & 6.0 & 6.31 \\
        ... & ... & ... & ... & ... & ... & ... & ... & ... & ... & ... & ... & ... & ... & ... & ... \\
    \enddata
    \tablecomments{This table is published in its entirety in a machine-readable format. A portion is shown here for guidance regarding its form and content.}
    \tablenotetext{a}{MYStIX MPCM ID number from \cite{Fe+13}. Only sources present in the MPCM catalog have an entry in this column.}
    \tablenotetext{b}{An entry of Y: source has JHKs excess, N: No excess.}
    \tablenotetext{c}{Isochronal mass determined through weighted averaging of discrete isochronal masses. Mass $\leq0.09$~M$_{\odot}$ (the lowest isochronal mass) indicated by -1; Mass $\geq100$~M$_{\odot}$ (the highest isochronal mass) indicated by 100; thus, no mass estimates can be provided for these sources. Mass values between $1.8-11$~M$_{\odot}$ correspond to the intersection with the isochrone with the smallest $A_V$. The uncertainty comes from an empirical relation developed using the statistical sampling method of \cite{Ro+26}.}
    \tablenotetext{d}{The uncertainty comes from an empirical relation developed using the statistical sampling method of \cite{Ro+26}.}
    \tablenotetext{e}{Determined via 2D projection using confirmed OB stars from \cite{St+24}. Equivalent units include $\rm erg~s^{-1}~cm^{-2}$ and $\rm 625~G_0$.}
\end{deluxetable*}

\section{Analysis} \label{sec:analysis}
\subsection{Mass and Extinction}
We used a 1~Myr PARSEC isochrone to estimate mass and extinction for each object from the $(J-H)$ versus $J$ and $(H-K_{\rm s})$ versus $H$ CMDs (referred to in this section as JCMD and HCMD, respectively). These give slightly different estimates for mass, with the distribution for the MYStIX-selected sources shown in Figure \ref{fig:Masses}. These objects cover the full mass range of the PARSEC isochrone below 1.8~M$_{\odot}$ (i.e., $\geq0.09$~M$_{\odot}$), though most sources we find are of roughly solar mass.

Using the same isochrone, we calculated the extinction towards each source (Section \ref{sec:dereddening}). The results are shown in Figure \ref{fig:Extinction}. The extinction measurement suffers from the same degeneracy as the mass estimate, so we only consider sources $<1.8$~M$_{\odot}$. Sources with negative extinction lie to the left of the isochrone and are likely the result of photometric errors or the accidental association of MYStIX objects with older foreground sources. As a rough approximation of the accidental association rate, we find that 43 (5\%) of the MYStIX-selected sources have $(J-H)<1$, which is more consistent with the foreground population (see Figure \ref{fig:isochrones}). These sources are roughly evenly distributed between the M17-SW dark region \citep[see][]{handbookM17} and the open central region, but do not constitute a significant fraction of sources in either region.

The median/mean extinction for sources with masses $<1.8$~M$_{\odot}$ is $A_{V}=6.5/7.1$~mag and $A_{V}=10/11$~mag for JCMD and HCMD, respectively. Using JCMD, we calculated most sources to have $A_{V}$ between $\sim$3 and 10~mag. This is a broad range, but it is reasonable considering the qualitative difference between the dense clouds to the north/southwest and the open region in the center/southeast that illustrates the complexity of the region. JCMD and HCMD give different estimates for the mass and extinction because any K-band excess influences the color on HCMD. For this reason, in Table \ref{tab:Data} and in our analysis, we use the mass and extinction estimates determined using JCMD. We explore the effects of changing the age estimate by adopting 0.3 and 3~Myr isochrones in Appendix \ref{app:Isochrones}.

\subsection{NIR Excess} \label{sec:IR excess}
A source is determined to have a NIR excess if the measured NIR flux is higher than predicted from the stellar photosphere. For YSOs, excess is observed at $\gtrsim2~\unslant\mu$m and usually comes from hot ($\gtrsim1000$~K) dust in the inner disk.

The reddening slope describes the relative reddening between $(J-H)$ and $(H-K_{\rm s})$: ${\rm E}(J-H)/{\rm E}(H-K_{\rm s})$ and is equivalent to the slope of the reddening vector on a CCD. Using this equation and the median $(J-H)$ and $(H-K_{\rm s})$ colors of the MYStIX-selected probable cluster members, we determined the reddening slope for the $(J-H)$ versus $(H-K_{\rm s})$ diagram to be $1.74\pm0.14$, in agreement with the results of \cite{RL85}, \cite{Ca+89}, \cite{Me+97}, and \cite{In+05}. As the value of 1.73 from \cite{In+05} used by P11 falls within our uncertainty, we adopt their value for consistency. Following P11, a source has JHKs excess if it lies at least 1$\unslant\sigma_{\rm phot}$ and more than 0.05~mag to the right and below the reddening slope and above $(J-H)=0$. Sources in this region have dereddened colors more consistent with that of the classical T Tauri star (CTTS) or Herbig Ae/Be loci than the main sequence (see Figure \ref{fig:populations}). We find 2042~sources in the FOV with JHKs excess, of which 1082 have photometric uncertainties $<0.1$~mag.

To determine the disk fraction for the entire sample of 10,399 sources, we simply divided the number of sources with infrared excess by the total number of sources. We find an overall JHKs excess fraction of $2042/10339=19.8\pm0.5\%$ and $1082/4727=22.9\pm0.8\%$ for all sources with $<0.1$~mag photometric uncertainties. Uncertainties on the JHKs excess fractions were calculated through $\sqrt{N}$. These values do not represent the true disk fraction of the cluster, as many of the included sources are fore/background objects.

To select YSOs, we catalog matched with the MYStIX MPCM catalog, which consists almost entirely of X-ray-detected cluster members. This is a strong selection criterion because X-ray detections are sensitive to the magnetically-active YSOs \citep{Pa+81}, while insensitive to the field contaminants that are mostly not expected to be strong X-ray emitters. While an X-ray detection alone would not remove all contaminants, see Section \ref{sec:member selection}.

\begin{figure}
    \centering
    \includegraphics[width=0.473\textwidth]{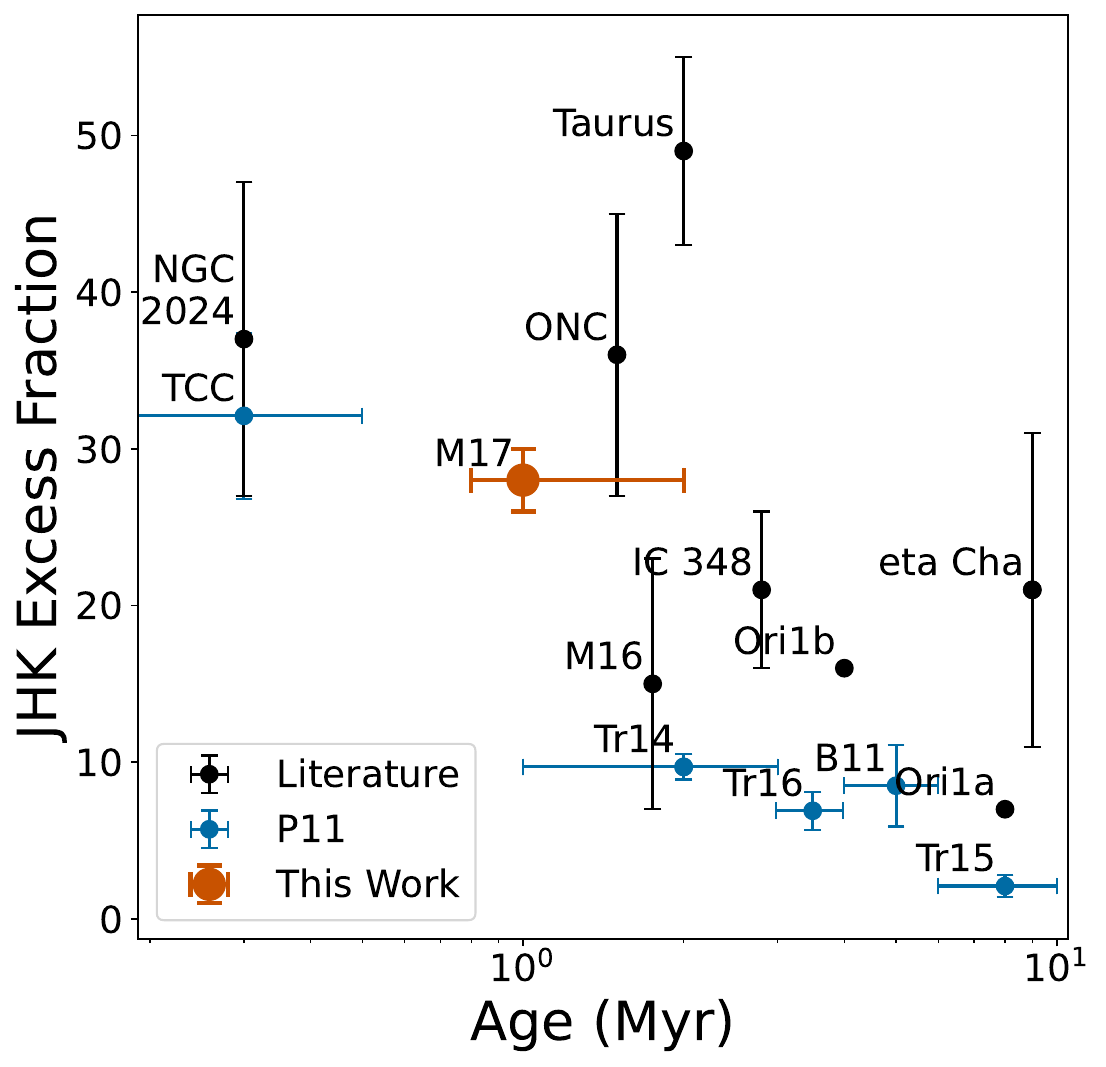}
    \caption{Disk fractions for 14 clusters within $\lesssim2.5$~kpc adapted from Figure 6 in P11. Points in blue from P11, points in black retrieved by P11 from the literature, and this work in dark orange. Note: The position of the point for NGC~2024 has been adjusted from its position in Figure 6 of P11, as \cite{Ha+00} state that the disk fraction estimate for the center of the cluster is not reliable. Therefore, we remove those sources and recalculate the disk fraction to be $37\pm10\%$.}
    \label{fig:Preibisch Disk Fractions}
\end{figure}

Because NIR light is reemitted mostly from the hotter inner portion of the disk, a NIR excess measurement at the location of an X-ray source in MYStIX implies the existence of a hot inner disk around a YSO. We measured a JHKs excess fraction of $277/1016=27\pm2\%$ for all MYStIX-selected sources and $261/932=28\pm2\%$ for those with photometric uncertainties $<0.1$~mag. The larger fraction compared to all sources is expected, as the majority of contaminants in the FOV are older, disk-less field stars, which artificially drives down the measured disk fraction, though there may also be some contaminants with NIR excess such as AGB stars, Be stars \citep{Re+10}, or star-forming galaxies \citep{La+07}. The comparison between the excess fractions of M17 and the regions in P11 is shown in Figure \ref{fig:Preibisch Disk Fractions}.

\subsection{Disk Fraction Versus UV within M17} \label{sec:DF v. UV}
\begin{figure*}
    \centering
    \includegraphics[width=\textwidth]{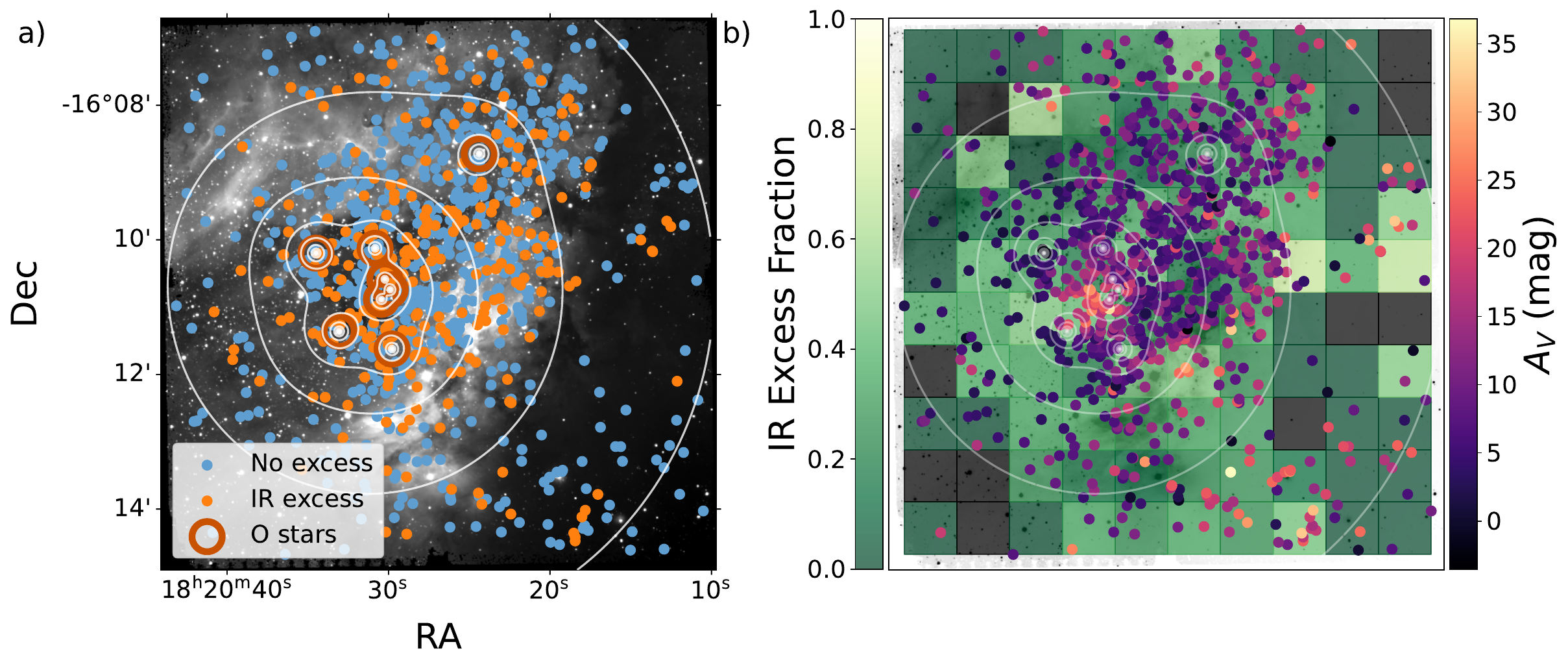}
    \caption{a) Ks-band image of M17. Colored dots represent MYStIX-selected sources. Orange dots have NIR excess, blue dots have no excess. Positions of O star cluster members confirmed by \cite{St+24} are circled in dark orange. Contours spaced 0.5~log(G$_{0}$) apart increase towards the center and represent the level of incident UV radiation. Outermost contour at ${\rm log(G_0)=3.5}$. b) Inverted background image with the same contours as (a). Grid represents the spatial dependence of disk fraction (left colorbar), where black squares contain no sources. Dot color (right colorbar) represents extinction towards each source. Note the increased extinction towards the IR-dark cloud at the center of the region, M17-SW on the right-hand side, and the irradiated clouds around the open central region.}
    \label{fig:X-ray Excess and UV}
\end{figure*}

The level of incident UV radiation that YSOs receive in M17 varies by more than three orders of magnitude as a function of position (log(G$_{0})=~\sim$$3.3-7.0$, see Figure \ref{fig:X-ray Excess and UV}). This makes the region a prime location to investigate the relationship between JHKs excess fraction and the level of incident UV radiation. However, because there is an unequal distribution of sources as a function of incident UV radiation in physical space, we have distributed the sources into bins with equal numbers of sources and ordered them by increasing incident UV radiation.

The IR-dark cloud at the center the FOV (see Figure \ref{fig:3-color}) almost completely covers the region of the cloud with log(G$_{0})>5$, creating two distinct regimes in the calculation of the disk fraction (see Figure \ref{fig:DF vs. UV}; the dashed line). Figure \ref{fig:DF vs. UV}a shows the trend between JHKs excess fraction and UV field strength for a choice of 25 bins of $\sim$37 sources each. The horizontal error bars represent the bin width, and the vertical error bars were calculated via $\sqrt{N}$. When accounting only for bins with log(G$_{0})<5$, we measure the slope of the correlation to be $0.07\pm0.02$ for any choice of total bin number $6\leq N_{\rm bins}\leq 37$ (bin size $25\leq N_{\rm sources}\leq 155$). The Pearson correlation coefficient represents the scatter in the correlation, with $+1/-1$ being perfectly correlated/anticorrelated. When applied to our data, the correlation tends to decrease with bin size, which we believe results from larger bins averaging more completely over the varying nebulosity of the region.

For a choice of 25 bins, as in Figure \ref{fig:DF vs. UV}, we find a slope of $0.084\pm0.002$ and a coefficient of 0.4, with a p-value of 0.1, for JHKs excess fraction versus UV. The slope of the trend is also nearly identical for both high-mass ($\geq1.8~{\rm M}_{\odot}$) and low-mass ($<1.8~{\rm M}_{\odot}$) sources, and a positive trend is observed for both the MYStIX-selected cluster members and for all sources. This is simultaneous with a negative trend in extinction versus UV for the MYStIX-selected sources with log(G$_0)<5$ (Figure \ref{fig:DF vs. UV}b) with a slope of $-5.7\pm0.7$ and a correlation coefficient of $-0.9$ with a p-value of $2.4\cdot10^{-7}$. These two trends result in a negative trend of JHKs excess fraction with extinction (Figure \ref{fig:DF vs. UV}c) with a slope of $-0.05\pm0.04$ and a correlation coefficient of $-0.5$ with a p-value of 0.03 for a choice of 25 bins.

The Pearson correlation coefficients and p-values vary with bin size, so the point biserial correlation coefficient is a better metric. This metric also varies between $-1$ and $+1$; however, it describes the correlation between a binary variable (in this case JHKs excess) and a continuous variable (UV flux or extinction). For JHKs excess and UV flux, we find a correlation coefficient of 0.07 with a p-value of $5\cdot10^{-6}$, and for JHKs excess and extinction, we find a coefficient of -0.05 with a p-value of $6\cdot10^{-4}$, indicating that the correlations between JHKs excess and both incident UV flux and extinction are statistically significant.

\section{Discussion} \label{subsec:disc}
Our JHKs photometric survey towards M17 is sensitive to $\lesssim21$~mag in the J-band. For MYStIX-selected cluster members, we estimate mass, with the distribution shown in Figure \ref{fig:Masses}, and extinction, with most sources $<1.8$~M$_{\odot}$ having $A_{V}$ between $\sim$3 and 10~mag and a median/mean $A_{V}$ of $6.5/7.1$~mag (Figure \ref{fig:Extinction}). We derive a disk fraction of $28\pm2\%$ for MYStIX-selected sources with $<0.1$~mag photometric uncertainties. Below, we discuss these results in comparison to previous surveys of M17 and other SFRs of similar age to explore the dependence of disk fraction on UV radiation. Within M17, we initially find a positive correlation between disk fraction and level of incident UV radiation, and we discuss some possible causes of this phenomenon.

\subsection{Comparison to Previous NIR Surveys of M17}
\label{sec:previous work}
Our measured extinction distribution is consistent with \cite{Br+07} and \cite{Ba+24}, who estimate $A_{V}$ of $3-15$~mag and $3.6-10.6$~mag, respectively. \cite{Ra+17} find $A_{V}$ of $5.4-13.6$~mag for OB stars and massive pre-main sequence stars, which can be seen through more extincting material than the low-mass sources we present, creating a systematic offset.

\cite{Ji+02}, using an $\lesssim18.7$~mag in J-band, $1.2^{\prime\prime}$ resolution, $\sim$$14^{\prime}\times14^{\prime}$ FOV IRSF/SIRIUS JHK survey of M17 report a JHK excess fraction of $\sim$$30\%$. This is consistent with the value we report, though field source removal was handled statistically via field observations in that work, instead of X-ray detections.

\cite{Br+07}, using the same IRSF/SIRIUS data, $17^{\prime}\times 17^{\prime}$ \textit{Chandra} observations, and 2MASS JHK observations for the portion of the \textit{Chandra} field not covered by the SIRIUS survey, found a JHK excess fraction in M17 of $\sim$$100/609=$$\sim$$16\%$ for all high-quality detections and $\sim$$12\%$ for those with masses $\geq2.0$~M$_{\odot}$. The discrepancy between this value and the results of both this work and that of the JHK surveys in this section likely comes from the focus of their survey on the shorter-lived disks around higher-mass stars and the relatively low completeness of their \textit{Chandra} and NIR observations compared to the later full MYStIX catalog and the HAWK-I survey presented here.

\cite{Ho+08}, with JHKs observations from VLT/ISAAC with $0.6^{\prime\prime}$ resolution and $\lesssim18.9$~mag J-band sensitivity in a $6.8^{\prime}\times8.8^{\prime}$ FOV toward M17 find a JHKs excess fraction of $\sim$$29\%$. This is consistent with our result, but was found using statistical star removal similar to \cite{Ji+02}. In summary, while the disk fraction presented in this work is consistent with those calculated using statistical star removal, we find a significantly greater disk fraction than the previous X-ray selected measurement, likely due to the shallower depths of previous X-ray and NIR surveys.

\subsection{Untangling the Disk Lifetime Discrepancy} \label{sec:Disk Lifetime}
Disks decay over time as material accretes onto the host star, evaporates, or coalesces into planets. External environmental effects contribute to this decay but can only be fully sampled in regions at d~$\gtrsim2$~kpc. However, low-mass stars are difficult to detect at these large distances, so they are undersampled in disk lifetime studies of these regions. This causes a larger-than-physical spread in the NIR excess fraction measurements in clusters of similar age, which can be seen in the $\pm20\%$ excess fraction vertical scatter for the $1-2$~Myr regions in Figure \ref{fig:Preibisch Disk Fractions}. Therefore, attempting to fit all regions with a single function to determine a universal disk lifetime causes confusion. For example, see Figure 1 of \cite{Pf+22}, who came to the conclusion that the discrepancy between the $5-10$~Myr disk lifetimes in nearby ($<200$~pc) clusters, and the $1-3$~Myr disk lifetimes in distant ($\ge200$~pc) clusters is mainly due to differences in limiting magnitude between observations. Observations of distant clusters tend to suffer from crowding and lower limiting absolute magnitudes, reducing the frequency of observed low-mass stars. High-mass stars tend to lose their disks faster \citep[e.g.,][]{Ca+06, Fa+12, Ya+14, Ri+15, Ri+18}, causing surveys that include more distant regions to report artificially short disk lifetimes.

For example, \cite{St+04} found a 27\% L-band excess fraction in NGC~3603, a massive $\sim$1~Myr-old, $\rm{d}=6.0\pm0.3$~kpc, starburst cluster. While they achieve 0.1~M$_{\odot}$ sensitivity in their JHK observations, photometric uncertainties due to crowding prohibited them from making NIR excess fraction estimates from their JHK photometry alone. Later, \cite{St+10} presented an L-band excess fraction of $6\pm2\%$ for the massive $\sim$2.5~Myr-old, $\rm{d}\sim8.5$~kpc \citep{Na+04} Arches cluster, located near the galactic center. However, that work was only sensitive down to $\sim$5~M$_{\odot}$, as opposed to the $\ll1~{\rm M}_{\odot}$ sensitivity of studies in nearby regions such as Taurus \citep[e.g.,][]{Lu18}. In general, studies of more distant regions ($\gtrsim4$~kpc) do not fully sample low-mass stars and are thus not directly comparable to studies of high-mass regions within $\sim$2~kpc.

However, there is significant evidence \citep[e.g., P11;][]{Gu+16, Co+22, Co+23, Ma+25} that the environment, and specifically external photovaporation, also influences disk lifetime. It is not ideal to directly compare the full disk fraction measurements in different regions, as the different ranges of observed stellar masses make the data sets incompatible. For example, because Taurus lies more than 10 times closer to Earth than M17 does, surveys of Taurus usually reach deeper limiting magnitudes. Based on the cluster membership of \cite{Lu18}, 2MASS photometry, and PARSEC isochrone analysis, we calculated that more than 20\% of Taurus cluster members have masses $\leq0.09~{\rm M}_{\odot}$, whereas in the data for M17 presented here, only 5\% of the sources in the full catalog, and less than 1\% of the MYStiX-selected sources, have masses that low.

\begin{deluxetable*}{cccccc}
    \tabletypesize{\small}
    \tablecaption{Summary of Disk Fraction (DF) Measurements\label{tab:Excess Fractions}}
    \tablehead{
    \colhead{SFR} & Central UV Field [log(G$_0$)] & \colhead{P11 DF} & \colhead{Default DF} & \colhead{Mass-limited DF} & \colhead{Ref.}
    }
    \startdata
    Taurus & $\sim0-2$   & $49\pm6\%$    & $105/380=28\pm3\%$  & $17/44=40\pm10\%$ & a \\
    ONC    & $\sim3-6$   & $36\pm9\%$    & $391/1333=29\pm1\%$ & $45/140=32\pm6\%$ & a, b \\
    M17    & $\sim4.5-7$ & N/A           & $261/932=28\pm2\%$  & $62/252=25\pm4\%$ & c \\
    Tr~14  & $\sim6-8$   & $9.7\pm0.8\%$ & $126/1210=10\pm1\%$ & $26/271=10\pm2\%$ & d, e \\
    \enddata
    \tablecomments{P11 DF refers to fractions from P11, Default DF from our analysis of all observed cluster members in each region, and mass-limited DF from the same data, limited to sources $0.8-1.8~{\rm M}_{\odot}$.}
    \tablerefs{a) \cite{An+25}, b) \cite{So+11}, c) \cite{St+24}, d) \cite{Sm06}, e) \cite{Be+23}
    }
\end{deluxetable*}

To isolate the environmental effects on the disk fraction, we correct for variable limiting magnitude by only considering sources in a common mass range across the subset of the P11 cluster sample within the age uncertainty of M17: Taurus, the ONC, and Tr~14\footnote{The UKIDSS JHK photometry available for M16 in MYStIX is not deep enough to provide a consistent comparison.}. To quantify the effect of setting a consistent mass-limit, we repeated our M17 disk fraction analysis for the three regions, also repeating the non-mass-limited (``default'') disk fraction calculation for Tr~14 to ensure that our measurement matched the one presented in P11. We used 2MASS photometry for the sources in the membership catalog of \cite{Lu18} for Taurus. For the ONC, we used MYStIX for membership selection and photometry. And for Tr~14, we used the catalog of P11 for photometry, combined with the catalog of \cite{Fe+11} to match the membership selection of P11.

We can directly compare regions only where our different datasets equivalently sample the IMF. For simplicity and to avoid the degenerate region of the 1 Myr isochrone, we limit our analysis to sources $<1.8~{\rm M}_{\odot}$. To determine the region where the different surveys equivalently sample the IMF, we make histograms of the masses of the sources in each region and take the ratio between them to see where they diverge. The ratio is relatively constant above 0.8~M$_{\odot}$ across all regions. Therefore, we report the excess fraction for sources $0.8-1.8~{\rm M}_{\odot}$ in each of these regions in Table \ref{tab:Excess Fractions}. It is easy to see that the mass-limited excess fractions for Taurus and the ONC are lower than the values given in P11, reducing the overall vertical scatter. This is because of the less-frequent detection of low-mass stars in distant regions, so the disk fraction-reduction effect is more substantial for closer regions: Taurus is most affected, while Tr~14 is largely unaffected, leading to a smaller range of disk fractions between the regions.

However, as is shown in Table \ref{tab:Excess Fractions}, while the Taurus and ONC JHK excess fractions are lower than previous results, the Taurus excess fraction is still significantly greater than the excess fraction in the ONC. Both are higher than the excess fraction in M17, while the excess fraction in Tr~14 remains significantly lower than M17. Tr~14 has a UV field of $\sim$$10^6-10^8$~G$_0$ at its center from $\sim$20 O stars in the cluster as well as external radiation from other nearby clusters to the northeast and southeast \citep{Sm06, Po+11}. This strips away the disks more rapidly than in M17, which experiences $\sim$$10^{4.5}-10^7$~G$_{0}$ in its center, leaving fewer YSOs with disks, even though Tr~14 is of a similar age to M17. Conversely, the ONC has only one dominant O7V and one O9.5V star, so nearby YSOs experience $\sim$$10^3-10^6$~G$_{0}$ \citep{Wi+19, An+25}, much less compared to M17 or Tr~14. Taurus is even more extreme, experiencing three orders of magnitude less UV radiation than the ONC \citep{An+25}. Therefore, we posit that different limiting magnitudes/mass sensitivity alone cannot explain the observed disk fraction disparity, and this remaining difference is likely partially a result of the different UV environments.

\subsection{Disk Fraction Versus UV within M17} \label{sec:DF v. UV Discussion}
Within M17, the observed positive trend of disk fraction with incident UV radiation (Figure \ref{fig:DF vs. UV}) is, at first, surprising, as we expect lower disk fractions in higher UV environments. However, we present here some possible mechanisms, both physical and observational, that could bias the measured disk fraction and force a positive correlation.

As was first shown by \cite{Ha+01} and is clear from Figure \ref{fig:Preibisch Disk Fractions}, older populations have systematically lower disk fractions. Assuming a monolithic collapse scenario, and considering that NGC~6618 is centrally concentrated in the M17 H\,\textsc{ii} region, it is reasonable to assume that the youngest YSOs might be closest to the center, with older YSOs having had more time to be ejected and/or migrate towards the edge of the region, \citep[][]{My12, Pi+13, Ge2+14, Ge+18}. This would lead to an outward age gradient and artificially increase the slope of the trend. This hypothesis is supported by the overall conclusion of \cite{Ge+14} that widely distributed populations are usually older than the main clusters. However, they observed no such population, nor an outward age gradient, in M17.

Models of SFR evolution \citep[e.g.,][]{Gu+22, Mi+23} show that existing YSOs remain relatively stationary for a few million years as feedback (e.g., from massive stars) disperses the gas, while new YSOs continue to form in the densest gas. Therefore, one might expect the youngest YSOs in the dense clouds to the north/south west and older YSOs in the open center of the region, which would instead cause a shallower or more negative correlation.

The dusty material in M17 makes it a region of spatially varying extinction, which has a threefold effect on measurements of the disk fraction versus UV radiation. First, material between the star and the observer might follow an alternate or spatially varying extinction law, which can change the reddening slope and make a source appear to have an infrared excess even if it does not have a disk \citep{CW98}. This, combined with the possibility of a spatially varying $R_V$ \citep[see][who find $R_V$ to range from $3.3-4.7$]{Ra+17}, could cause confusion.

The material can also absorb UV radiation, providing an effective shield between YSOs and incident UV radiation from OB stars, protecting the disks and prolonging their lifetimes \citep[e.g.,][]{Ad+04, Da+05, Al+19, Qi+22, Wi+23}. As nearly all of the OB stars are in the central, open region of M17, sources near or within the north or southwest clouds might experience some degree of shielding. This could artificially inflate the disk fraction, obscuring the true underlying relation between disk fraction and UV radiation. The IR-dark cloud lies in the highest UV regime, to the right of the dashed line in Figure \ref{fig:DF vs. UV}, so shielding would likely be most significant there, where otherwise one might expect the smallest disk fractions. However, ALMA/ACA data show a strong gas detection at the location of the IR-dark cloud at a significantly different velocity than M17, suggesting it lies in the foreground and thus does not shield the YSOs in the center of the region (R. Harrison, in preparation). We estimate an average extinction of $A_{V}=15.1$~mag for the dark cloud by measuring the average extinction of the YSOs affected by the cloud. \cite{An+25} test the effect of shielding in many clusters, and find a negligible ($\lesssim1\%$) total FUV flux decrease from intra-star extinction in the highest-UV environments they study, with FUV-reduction only significant in the lowest-UV environments.

Finally, if sufficient material exists between a source and the observer, the extinction can obscure some sources altogether, leading to uneven survey depth across the region. There is clear evidence that this is occurring in M17. The full catalog shows a drastic drop in source density at the edge of the large clouds in the north and southwest of the FOV, with many more sources in the open regions in the center and east/southeast of the FOV. See also Figure \ref{fig:DF vs. UV}, which shows that the range of UV fluxes that coincides with the positive NIR excess correlation (${\rm log(G_0)\lesssim5}$) also coincides with a negative correlation between UV flux and extinction. This $\sim$10~mag decrease in extinction from the edge of the region means that there is a greater detection rate of low-mass stars in the evacuated central region than in the clouds. For example, at $A_V=10$~mag, we detect sources down to $\sim$0.2~M$_{\odot}$, whereas at $A_V=15$~mag we detect sources down to only $\sim$0.7~M$_{\odot}$.

The exact dependence of disk lifetime on host star mass is still unknown, though one estimate comes from \cite{Ya+14}, who determined disk lifetime (calculated from JHK excess) for YSOs $\leq1$~M$_{\odot}$ to vary with stellar mass as $M_*^{-0.7\pm0.3}$. If this is true, it would mean that 1~M$_{\odot}$ YSOs lose their disks roughly $2-5$~times faster than 0.2~M$_{\odot}$ YSOs. We believe the measured change in extinction (and thus mass) to be the predominant cause of the observed positive JHKs excess fraction correlation.

\begin{figure*}
    \centering
    \includegraphics[width=0.8\textwidth]{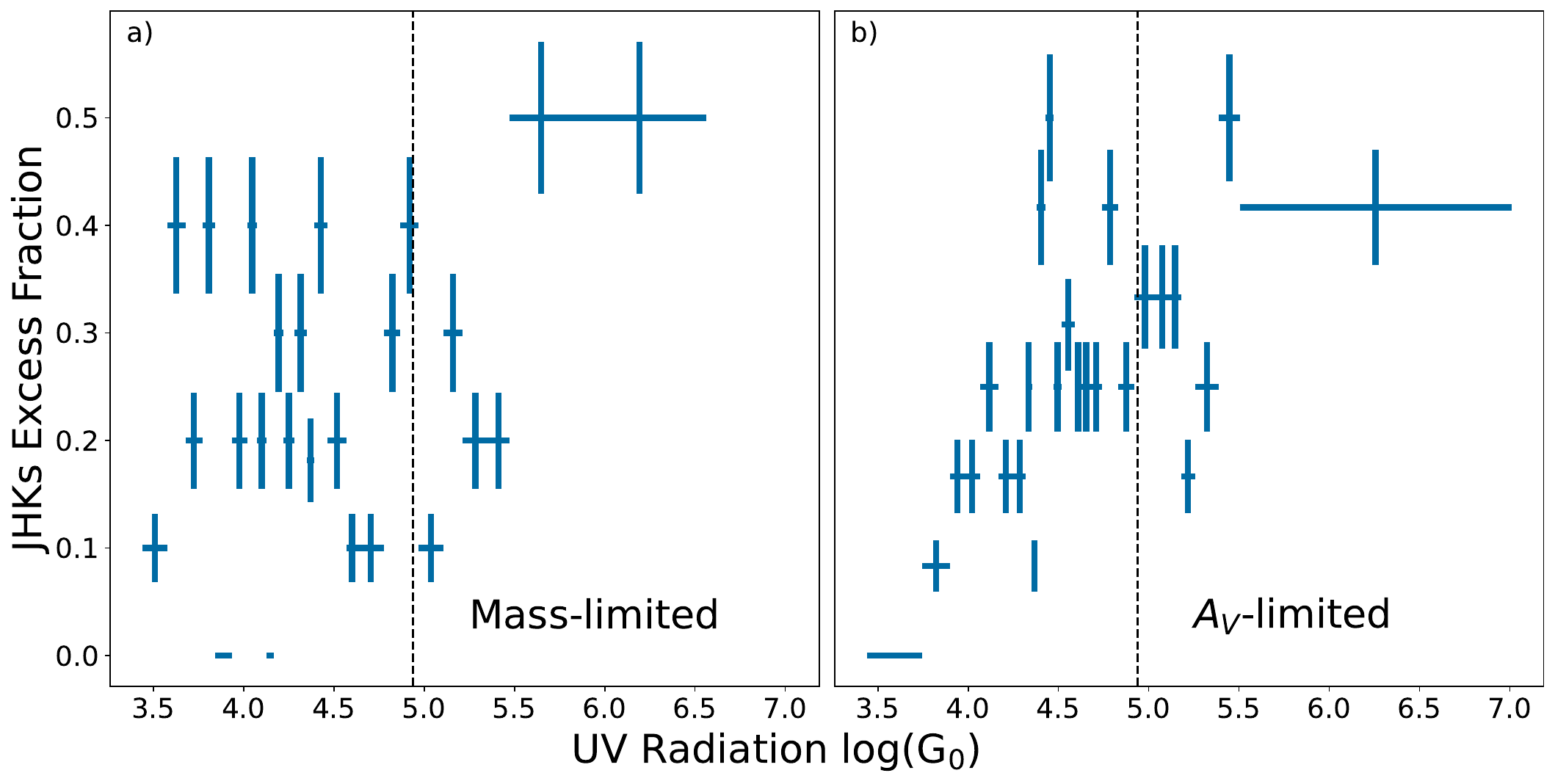}
    \caption{JHKs excess fraction for sources a) with masses $0.8-1.8$~M$_{\odot}$ and b) with $A_V<6.5$~mag (the median value) and $<1.8$~M$_{\odot}$. Panel (a) shows no correlation, implying that even with extinction allowed to vary, it has no remaining effect on the JHKs excess fraction when considering this range of masses. However, (b) shows a weak positive correlation, implying the limited extinction still has an impact on the masses of sources that are observed. However, removal of the highest- and lowest-UV points leaves at most a tentative correlation. Horizontal error bars denote bin width, vertical error bars from $\sqrt{N}$.}
    \label{fig:mass-limited DF vs. UV}
\end{figure*}

Figure \ref{fig:mass-limited DF vs. UV} shows the disk fraction for the $0.8-1.8$~M$_{\odot}$ and $A_V<6.5$~mag, $<1.8$~M$_{\odot}$ subsamples as a function of UV. Figure \ref{fig:mass-limited DF vs. UV}a shows that when the mass-limit is imposed and the effects of mass are removed, there is no correlation between disk fraction and the local UV field. This implies that even with extinction allowed to vary, it has little to no remaining effect on the observed sources and thus the disk fraction when considering this range of masses. However, Figure \ref{fig:mass-limited DF vs. UV}b shows that when only the extinction is limited, a weak positive correlation remains, showing that extinction plays a reduced but non-negligible role in the masses of observed sources for the applied extinction limit. These results imply that extinction is largely responsible for the unexpected positive correlation seen in Figure \ref{fig:DF vs. UV}, while the true underlying relation is uncorrelated.

This is expected, as we would not necessarily predict any correlation in dense, dynamically-evolved regions. While M17 is young \citep[$\sim$1~Myr;][]{handbookM17}, dynamical mixing (the random motion of gas, dust, YSOs, etc. throughout a region due to gravitational interactions) over 1~Myr is significant, and this, combined with 2D projection effects, can obscure how UV radiation affects the disk lifetime. \cite{Pa+21}, with N-body dynamical evolution simulations of external photoevaporation in relatively low-density SFRs ($440-630$~M$_{\odot}$ within a 1 or 5~pc radius), find no relationship between disk mass and distance from the massive star(s) in their simulations by 1~Myr, suggesting that each YSO has had a relatively equal exposure to UV radiation. We expect this effect to be even stronger in a higher-density environment such as M17.

\subsection{Spatial Distribution of the Disk Fraction}
Many authors studying the spatial variation of the disk fraction with UV, using some combination of JHK and \textit{Spitzer} $3.6-24$~$\unslant\mu$m observations of distant ($\gtrsim1-2$~kpc) SFRs have found evidence for a correlation. \cite{Ba+07} claim that YSOs $<0.5$~pc from an O star in NGC~2244 ($2-3$~Myr) have a lower disk fraction ($27\pm11\%$) than those farther away ($45\pm6\%$, a $\sim$1.2$\unslant\sigma$ difference). However, there is no clear trend as distance increases, with uncertainties in each bin consistent with the $<0.5$~pc disk fraction. This is likely evidence of dynamical mixing, similar to what we find in M17 (see Section \ref{sec:DF v. UV Discussion} and Figure \ref{fig:mass-limited DF vs. UV}). \cite{Fa+12} observed a similar phenomenon in Pismis 24 ($\sim$1~Myr). For YSOs $>0.5$~M$_{\odot}$, they found a disk fraction of $\sim$19\% for YSOs $<0.6$~pc from the dominant UV source in the region, and $\sim$$36-38\%$ for those at all larger distances. They claim that the spatial distributions of stars with masses greater and lower than 1~M$_{\odot}$ are not significantly different, but they do not directly account for stellar mass when calculating the disk fraction. \cite{Gu+23} found the disk fraction of Cygnus OB2 ($2-3$~Myr) to decrease from $\sim$39\% to $\sim$18\% between log(G$_0)=3.0-4.5$. They also found a similar correlation using mass-limited samples. According to the authors, this region is ``dynamically not evolved,'' so the presence of a correlation is not surprising.

Conversely, \cite{Ro+11} found no systematic relationship between distance to either of the O stars and the presence of a disk for YSOs limited to $\sim$1~M$_{\odot}$ (their completeness limit) in the $3-5$~Myr-old IC 1795 OB association. \cite{Ri2+15} found no evidence for inner disk depletion for YSOs nearer to the O stars in four regions from the MYStIX survey, though they use ECDF distance analysis for disk-bearing and disk-free sources, in contrast to the other works referenced here. \cite{Da+24}, in their study of two star-forming clusters in the W5 region (with similar ages and distances of $<5$~Myr and $\sim$2~kpc, respectively), found the cluster with more overall FUV radiation to have a significantly lower disk fraction ($\sim$$17\pm1\%$ versus $\sim$$27\pm2\%$), though they found no significant disk fraction variation as a function of position within either cluster when including all sources above their completeness limit of $>0.1$~M$_{\odot}$.

Together, this suggests that for individual, dynamically evolved regions, the current spatial variation of the UV field has little to no impact on the disk fraction and thus the disk lifetime.

\section{Conclusions} \label{sec:conc}
We present the deepest (${\rm J}<21$~mag) and highest spatial-resolution ($\sim\!600-1000$~AU) JHKs survey of M17, revealing 10,339~sources in an $\sim$$8^{\prime}\times8^{\prime}$ ($\sim$$4 \rm pc \times 4 \rm pc$)~FOV using VLT/HAWK-I. The extreme nebulosity in M17 demanded the development of a bespoke mosaicking method. The location of M17 on the Galactic plane necessitated the use of the MYStIX MPCM catalog, consisting mostly of X-ray-active YSOs, to select cluster members from field contaminants and calibrate our photometry. Using a PARSEC 1~Myr isochrone and by dereddening sources along an \cite{In+05} $R_{V}=4$ reddening vector, we find sources down to $\leq0.09$~M$_{\odot}$, with most sources having roughly solar mass and an $A_{V}$ between $\sim$3 and 10~mag (median $A_{V}=6.5$~mag), though we emphasize that individual mass estimates between $1.8-11$~M$_{\odot}$ are degenerate and highly uncertain. We determined a JHKs excess fraction of $28\pm2\%$ for the MYStIX-selected probable cluster members. This result is consistent with other observations of M17 made using statistical contaminant removal, but is higher than the previous X-ray-selected sample, likely because of the shallower depth of previous studies. For the mass-limited ($0.8-1.8$~M$_{\odot}$) sample, we find no correlation between JHKs excess fraction and incident UV radiation within M17, likely due to dynamical mixing. However, comparing the overall JHKs excess fraction for the same mass limit between regions of similar age, we find a clear trend of decreasing JHK(s) excess fraction with increasing UV. The influence of dynamical mixing and observational biases on the disk fraction highlights the need for a homogeneous, systematic disk fraction study across a range of clusters. Applying this methodology across regions will allow for the quantification of the relationship between external photoevaporation by UV radiation and the disk lifetime.

\begin{acknowledgments}

The authors would like to thank the anonymous referee for the detailed and thoughtful review, Lodovico Coccato for assistance in developing the final HAWK-I data reduction method, and Rachel Harrison for the indication that the IR-dark cloud lies in the foreground. This paper is based on data obtained with ESO telescopes at the Paranal Observatory under programme ID 111.252G. 
S.~M. and M.~R. acknowledge support from NSF CAREER grant 2339164. T.~J.~H.  acknowledges a Dorothy Hodgkin Fellowship, UKRI guaranteed funding for a Horizon Europe ERC consolidator grant (EP/Y024710/1). D.~I. acknowledges support from collaborations and/or information exchange within NASA’s Nexus for Exoplanet System Science (NExSS) research coordination network sponsored by NASA’s Science Mission Directorate under agreement no. 80NSSC21K0593 for the programme ‘Alien Earths'. R.~J.~P. acknowledges support from the Royal Society in the form of a Dorothy Hodgkin Fellowship.
\end{acknowledgments}

\begin{contribution}



S.~M. led the investigation, methodology, formal analysis, visualization, and validation of the data, and was responsible for writing, reviewing, editing, and submitting the manuscript.

M.~R. was responsible for the initial research concept, funding acquisition, data acquisition, project administration, supervision, validation, and reviewing and editing the manuscript.

M.~A. was responsible for the initial research concept, data acquisition, project administration, supervision, validation, and reviewing and editing the manuscript.

T.~H., D.~I., A.~M., R.~J.~P., A.~W., and P.~Z. developed the initial research concept and reviewed the manuscript.

\end{contribution}

%
\facility{VLT:Yepun (HAWK-I)}

\software{aplpy \citep{aplpy},
          astrometry \citep{astrometry},
          astropy \citep{astropy:2013, astropy:2018, astropy:2022},
          astroquery \citep{astroquery},
          dust\_extinction \citep{dust_extinction},
          esorex \citep{EsoRex},
          matplotlib \citep{matplotlib},
          numpy \citep{Numpy},
          pandas \citep{Pandas},
          photutils \citep{photutils},
          reproject (https://reproject.readthedocs.io),
          SAOImageDS9 \citep{ds9},
          scipy \citep{Scipy},
          uncertainties (https://pythonhosted.org/uncertainties/)
          }


\appendix

\section{Effect of Variable Isochrone Age}
\label{app:Isochrones}

\begin{figure*}
    \centering
    \includegraphics[width=0.8\linewidth]{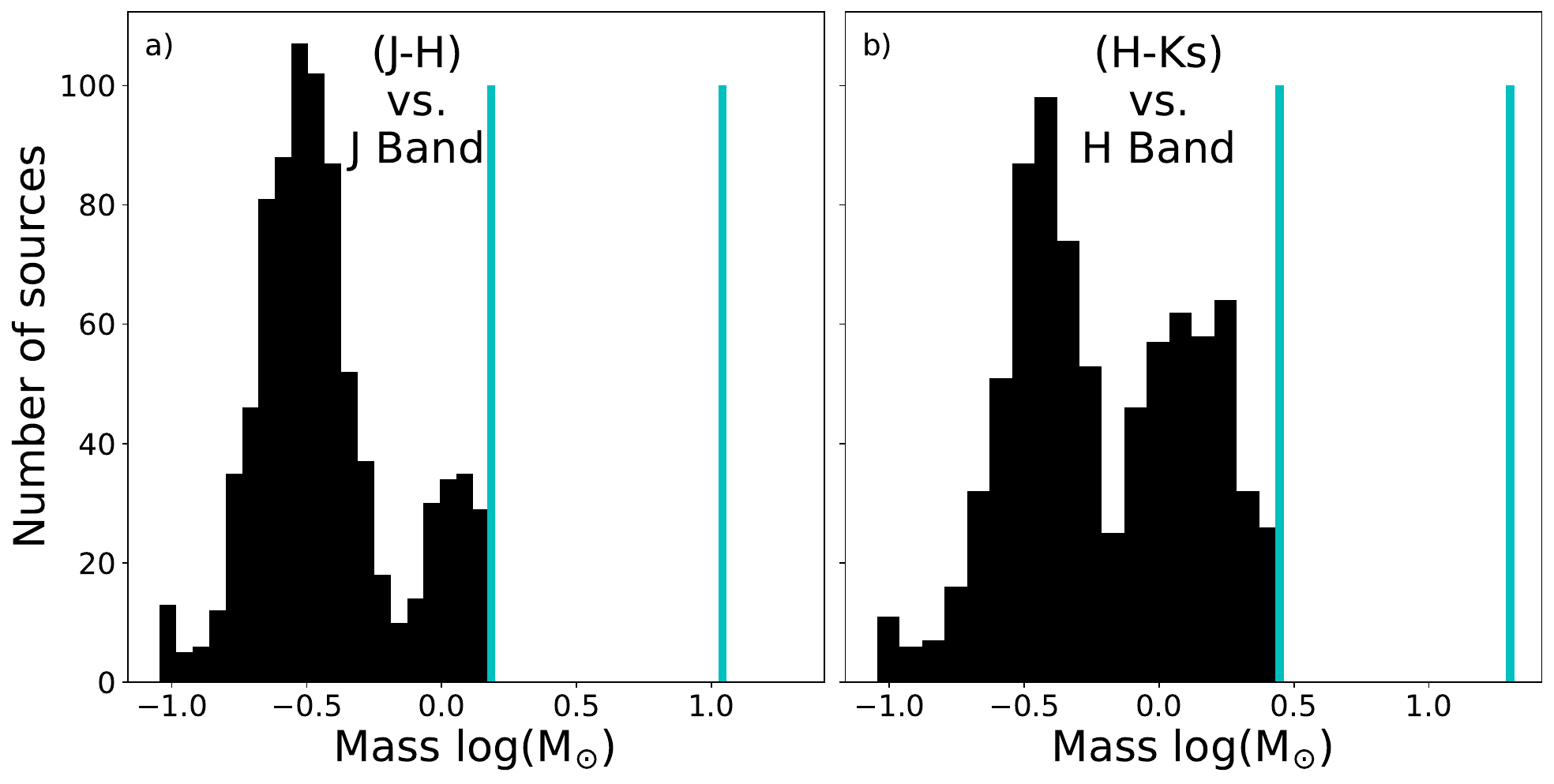}
    \caption{Mass measurement for MYStIX-selected sources in M17 at 0.3 Myr. Cf. Figure \ref{fig:Masses}.}
    \label{fig:Masses 300 kyr log}
\end{figure*}

\begin{figure*}
    \centering
    \includegraphics[width=0.8\linewidth]{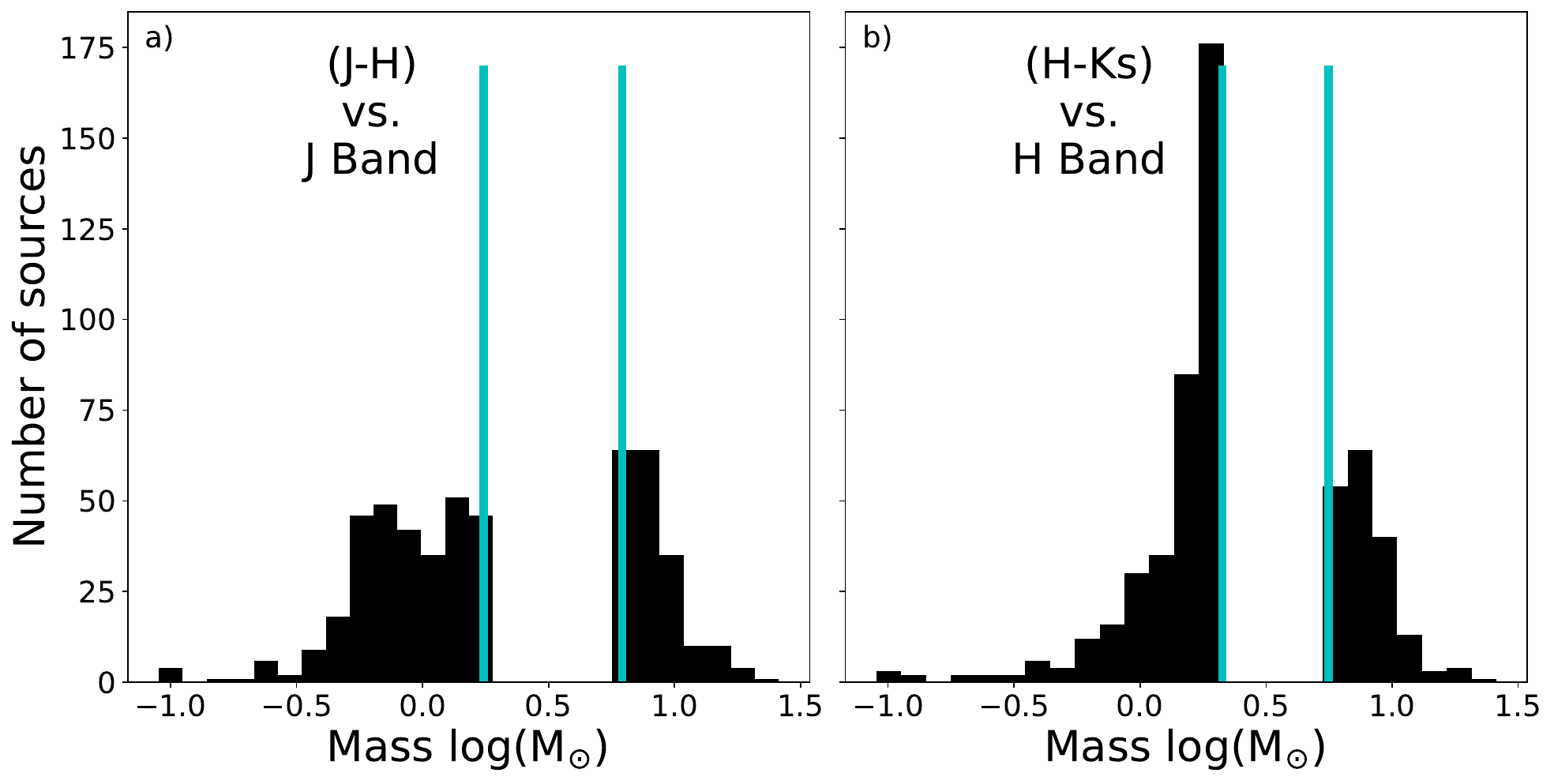}
    \caption{Mass measurement for MYStIX-selected sources in M17 at 3 Myr. Cf. Figure \ref{fig:Masses}.}
    \label{fig:Masses 3 Myr log}
\end{figure*}

\begin{figure*}
    \centering
    \includegraphics[width=0.8\linewidth]{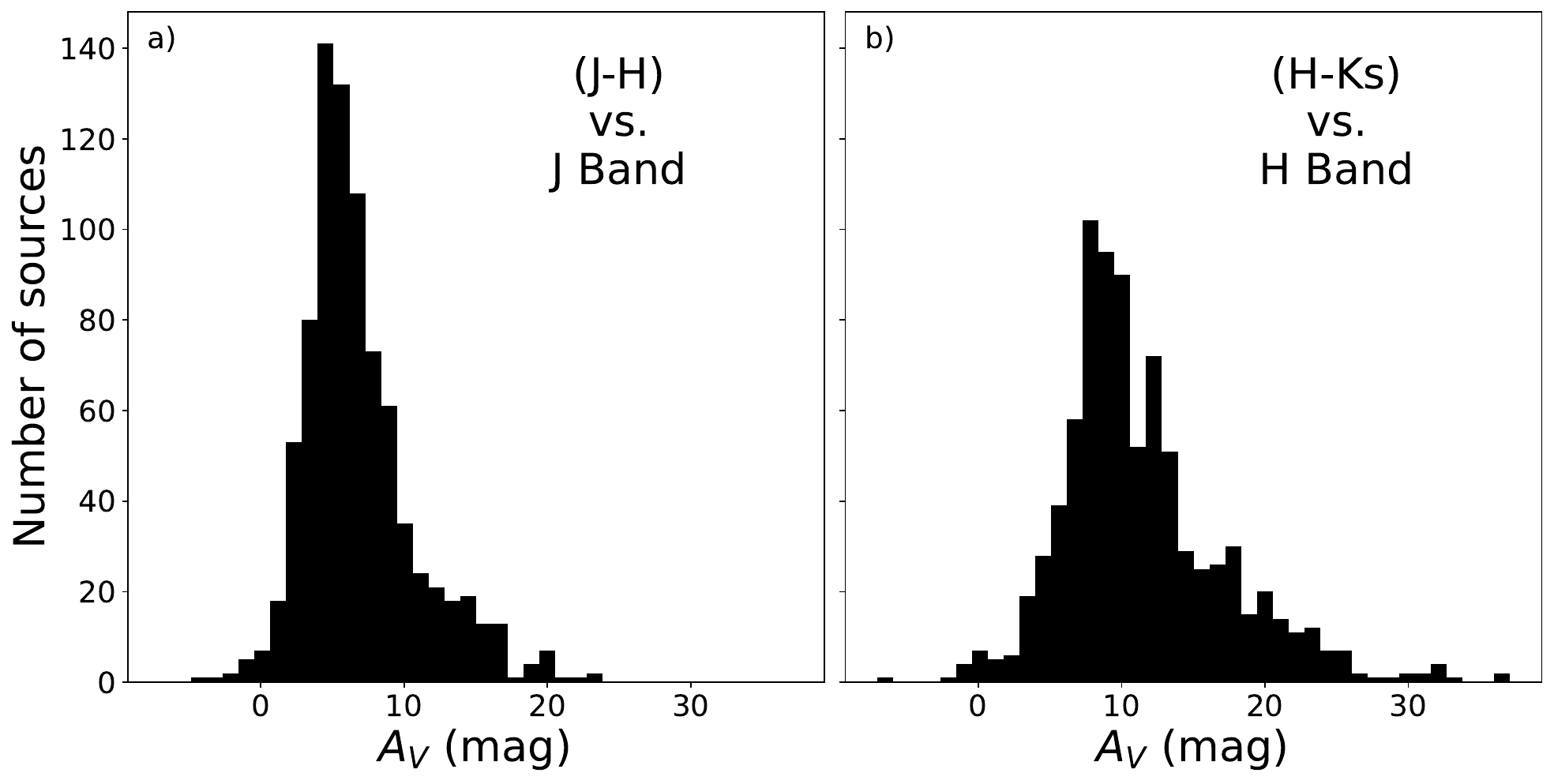}
    \caption{Extinction measurement for MYStIX-selected sources $<1.52$ M$_{\odot}$ in M17 at 0.3 Myr. Cf. Figure \ref{fig:Extinction}.}
    \label{fig:Extinction 300 kyr}
\end{figure*}

\begin{figure*}
    \centering
    \includegraphics[width=0.8\linewidth]{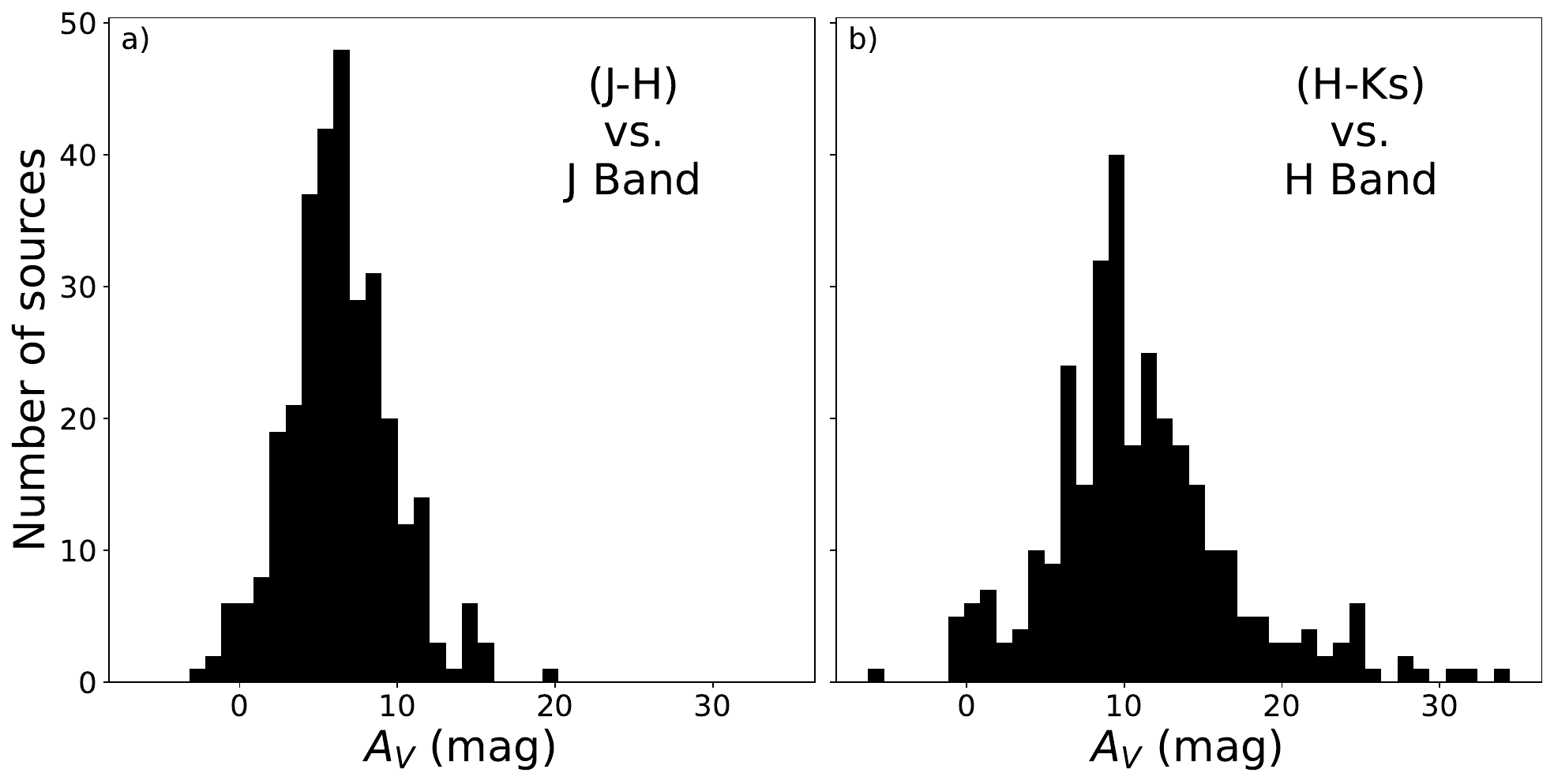}
    \caption{Extinction measurement for MYStIX-selected sources $<1.75$ M$_{\odot}$ in M17 at 3 Myr. Cf. Figure \ref{fig:Extinction}.}
    \label{fig:Extinction 3 Myr}
\end{figure*}

The measurement of mass and extinction depends on the age of the cluster. We demonstrate this dependence by repeating the analysis with 0.3 Myr and 3 Myr isochrones from the PARSEC models. The quantities that change are shown in the figures below. The change in mass measurements are shown in Figures \ref{fig:Masses 300 kyr log} \& \ref{fig:Masses 3 Myr log}, while the changes in extinction measurements are shown in Figures \ref{fig:Extinction 300 kyr} \& \ref{fig:Extinction 3 Myr}. Because the shape of the isochrone changes with age, the range of masses for which the dereddening process produces degeneracies in the J-band changes from $1.8-11$ M$_{\odot}$ to $1.52-23.1$ M$_{\odot}$ (0.3 Myr)/$1.75-6.4$~M$_{\odot}$ (3 Myr). We measured the disk fraction for sources $<1.52$ M$_{\odot}$ to be $27\pm2\%$ (0.3 Myr), $29\pm3\%$ (1 Myr), and $36\pm4\%$ (3 Myr).

\bibliography{main}{}
\bibliographystyle{aasjournalv7}



\end{document}